\documentclass[%
reprint,
superscriptaddress,
preprintnumbers,
amsmath,
amssymb,
floatfix,
]{revtex4-2}

\usepackage{graphicx}
\usepackage{dcolumn}
\usepackage{bm}
\usepackage[utf8]{inputenc}
\usepackage[T1]{fontenc}
\usepackage{mathptmx}
\usepackage{textcomp}
\usepackage{siunitx}
\usepackage{bigints}

\usepackage{hyperref}
\usepackage[switch]{lineno}
\usepackage[table,xcdraw,dvipsnames]{xcolor}
\usepackage{booktabs}
\usepackage{float}
\usepackage{color}
\usepackage{ulem}
\usepackage{multirow}

\begin{document}
	
\title{Approaching the Physical Limits of Specific Absorption Rate in Hyperthermia Applications}

\author{S.~Scheibler}
\affiliation{Magnetic \& Functional Thin Films Laboratory, Empa, Swiss Federal Laboratories for Materials Science and Technology, Ueberlandstrasse 129, 8600 Dübendorf, Switzerland}
\affiliation{Nanoparticle Systems Engineering Laboratory, Institute of Energy and Process Engineering (IEPE), Department of Mechanical and Process Engineering (D-MAVT), ETH Zurich, Sonneggstrasse 3, 8092 Zurich, Switzerland}
\affiliation{Particles-Biology Interactions Laboratory, Empa, Swiss Federal Laboratories for Materials Science and Technology, Ueberlandstrasse 129, 8600 Dübendorf, Switzerland}
\author{H.~Wei}
\affiliation{Fraunhofer IMTE, Fraunhofer Research Institution for Individualized and Cell-Based Medical Engineering, Mönkhofer Weg 239a, 23562 Lübeck, Germany}
\author{J.~Ackers}
\affiliation{Fraunhofer IMTE, Fraunhofer Research Institution for Individualized and Cell-Based Medical Engineering, Mönkhofer Weg 239a, 23562 Lübeck, Germany}
\author{S.~Helbig}
\affiliation{Physics of Functional Materials, Faculty of Physic, University of Vienna, Kolingasse 14-19, 1090 Vienna, Austria}
\affiliation{Research Platform MMM Mathematics – Magnetism – Materials, University of Vienna, Kolingasse 14-19, 1090 Vienna, Austria}
\author{S.~Koraltan}
\affiliation{Physics of Functional Materials, Faculty of Physic, University of Vienna, Kolingasse 14-19, 1090 Vienna, Austria}
\affiliation{Research Platform MMM Mathematics – Magnetism – Materials, University of Vienna, Kolingasse 14-19, 1090 Vienna, Austria}
\author{R.~Peremadathil-Pradeep}
\affiliation{Magnetic \& Functional Thin Films Laboratory, Empa, Swiss Federal Laboratories for Materials Science and Technology, Ueberlandstrasse 129, 8600 Dübendorf, Switzerland}
\affiliation{Department of Physics, University of Basel, Klingelbergstrasse 82, 4056 Basel, Switzerland}
\author{M.~Krupi{\'n}ski}
\affiliation{Institute of Nuclear Physics Polish Academy of Sciences, Radzikowskiego 152, 31-342 Krak{\'o}w, Poland}
\author{ M.~Graeser}
\affiliation{Institute of Medical Engineering, University of Luebeck, Ratzeburger Allee 160, 23562 Luebeck, Germany}
\affiliation{Fraunhofer IMTE, Fraunhofer Research Institution for Individualized and Cell-Based Medical Engineering, Mönkhofer Weg 239a, 23562 Lübeck, Germany}
\author{D.~Suess}
\affiliation{Physics of Functional Materials, Faculty of Physic, University of Vienna, Kolingasse 14-19, 1090 Vienna, Austria}
\affiliation{Research Platform MMM Mathematics – Magnetism – Materials, University of Vienna, Kolingasse 14-19, 1090 Vienna, Austria}
\author{I.~K.~Herrmann}
\affiliation{Nanoparticle Systems Engineering Laboratory, Institute of Energy and Process Engineering (IEPE), Department of Mechanical and Process Engineering (D-MAVT), ETH Zurich, Sonneggstrasse 3, 8092 Zurich, Switzerland}
\affiliation{Particles-Biology Interactions Laboratory, Empa, Swiss Federal Laboratories for Materials Science and Technology, Ueberlandstrasse 129, 8600 Dübendorf, Switzerland}
\affiliation{Ingenuity Lab, Balgrist University Hospital and University of Zurich, Forchstrasse 340,8008 Zürich, Switzerland}
\author{H.~J.~Hug}
\affiliation{Magnetic \& Functional Thin Films Laboratory, Empa, Swiss Federal Laboratories for Materials Science and Technology, Ueberlandstrasse 129, 8600 Dübendorf, Switzerland}
\affiliation{Department of Physics, University of Basel, Klingelbergstrasse 82, 4056 Basel, Switzerland}
\email{hans-josef.hug@empa.ch}

\begin{abstract}
	Magnetic nanoparticle-based hyperthermia has emerged as a promising therapeutic modality for treating malignant solid tumors that exhibit resistance to conventional cancer treatments, including chemotherapy and radiation. Despite the clinical approval of superparamagnetic iron oxide nanoparticles (SPIONs) for the adjunct treatment of recurrent glioblastoma, their therapeutic potential is undercut by chemical synthesis-inherent limitations such as low saturation magnetization, superparamagnetic characteristics, and a wide nanoparticle size distribution. As an alternative, synthetic antiferromagnetic disk particle (SAF-MDPs) designs have been explored. However, to date, their
	effectiveness is constrained by undesirable magnetic behaviors; the reported in-plane magnetized SAF-MDPs displayed hysteresis-free magnetization loops, and those with perpendicular magnetization encountered prohibitively high coercive fields. 
	Here, we introduce an micromagnetic modelling-based SAF-MDP design with in-plane magnetization, optimized through specific uniaxial anisotropy adjustments to avert the spin-flop phenomenon and eliminate hysteresis-free hard-axis magnetization loops, paired with a mechanofluidic modeling approach to assess the alignment of the SAF-MDP to the applied alternating magnetic field (AMF).
	Magnetic Force Microscopy characterization provides unprecedented insights into the particle switching behaviour on a single particle scale. This comprehensive strategy spanning micromagnetics and advanced magnetic characterization enables the design of particles with heating efficiencies to approach the theoretical maximum, dictated by the saturation magnetization of the utilized materials and limited solely by the biologically acceptable frequencies and amplitudes of the oscillating magnetic field. Our work not only addresses the limitations encountered by previous methodologies but also sets the stage for the development of advanced SAF-MDP designs and alignment techniques. This opens a new avenue to hyperthermia-based cancer therapy, delineated only by the boundaries of physical laws and biological safety standards.
	
\end{abstract}

\flushbottom
\maketitle

\newpage
\thispagestyle{empty}

\section*{Introduction}
Magnetic hyperthermia emerges a promising approach for the elimination of cancerous tissues. While it is generally regarded as a valuable addition to established treatment options such as surgery, radiotherapy, and chemotherapy, magnetic hyperthermia in its current form faces considerable constraints due to poor heating efficiency and unfavorable heat dissipation.. These limitations lead to drastic consequences, including the inability to eradicate small sized tumors. Additionally, the poor heating efficiencies of contemporary magnetic hyperthermia necessitate the delivery of large doses of nanoparticles to the tumors. This requirement further limits the range of tumors that can be effectively targeted and treated, as not all tumor sites can safely or feasibly receive such high concentrations of nanoparticles. Magnetic nanoparticle based hyperthermia typically employs superparamagnetic iron oxide nanoparticles (SPIONs). Such particles have been approved for the treatment of recurrent glioblastoma in Europe, and more recently, by the U.S. Food and Drug Administration (FDA) for clinical trials for the treatment of prostate and pancreatic cancer. Owing to the necessity for substantial doses of particles within the tumor tissue, these particles are commonly administered directly into the tumor, its surrounding vasculature, or the peritumoral area prior to the application of an alternating magnetic field (AMF). This leads to magnetic hysteresis losses within the particles and consequently to a locally increased temperature leading to tumor cell death. The SPIONs are typically manufactured bottom-up, by chemical synthesis, carefully tuning their size distribution such that the largest particles have a volume and with it an anisotropy energy barrier which can be overcome at room temperature within a short timescale to prevent agglomeration. This is to obtain superparamagnetic properties of the particles and with it a vanishing magnetic moment at zero applied field needed to minimize an attractive magnetic inter-particle interaction which would destabilize a particle suspension. For slowly alternating magnetic fields, for example, for the field sweeping time in a classical vibrating sample magnetometry (VSM) experiment, the $M$($H$)-loop of superparamagnetic particles is Langevin-like and thus does not show any coercivity and hysteresis\,\cite{Gavilan2021} (blue curve in Fig.\,\ref{Fig:limitations}{\textbf{a}}). 
 
However, because SPIONs exhibit effective magnetic anisotropy $K$ due to crystalline and/or shape anisotropy $K$, the $M(H)$-loop develops a finite hysteretic loss area $A$ (red curve and shaded area in Fig.\,\ref{Fig:limitations}{\textbf{a}}) increasing with the frequency $f$ of the applied AMF.  This relationship is elucidated by Sharrock's equation,\cite{Sharrock1994}, which demonstrates how the coercive field $H_{\rm c}$ increases as the duration of the experimental field pulse $t_0$ decreases, with $t_0$ being inversely proportional to $f$ as
\begin{equation}
H_{\rm c}(t_0) = H_{\rm a} \left\{ 1 - \left[ \frac{k_{\rm B}T}{KV} \ln \left( \frac{t_0}{\ln(2) \tau_0} \right) \right]^{n} \right\}\,\, ,
\label{eq:Sharock}
\end{equation}
where $T$ is the temperature,  
$\tau_0 = 1/f_0 \approx 10^{-13}-10^{-9}$ is one over the attempt frequency and $0.5<n<0.7$ depending on the angle of the applied field and the anisotropy axis, and $H_{\rm a} = 2K/(\mu_0 M_{\rm s})$
is the anisotropy field, with $K$ being the effective uniaxial anisotropy of the particle and $M_{\rm s}$ its saturation magnetization.

The hysteretic power loss per volume can be estimated\,\cite{Hergt2007} from linear response theory (LRT) as 
\begin{equation}
P/V = f\cdot A = \mu_0 \pi H^2 f \chi '' \,\, , 
\label{eq:LRT}
\end{equation}
where $A$ is the area enclosed by the $M(H)$-loop, and with the imaginary part of the susceptiblity $\chi '' = \chi_0 \cdot [2\pi f \tau/(1+2\pi f \tau)^2]$ describing the de-phasing of the magnetic moment of the magnetic nanoparticles (MNPs) with respect to the AMF.  $\chi_0 = \tfrac{ \mu_0 M_{\rm s}^2 V}{3k_{\rm B}T}$ 
is the static susceptibility depending on the MNP's saturation magnetization $M_{\rm s}$ and their volume $V$. The hysteretic loss area $A$ of the $M(H)$-loop (red curve and shaded area in Fig.\,\ref{Fig:limitations}\textbf{a})  reaches a maximum for a frequency $f$ related to the total relaxation time $\tau$ as $2\pi f \tau = 1$. The total relaxation time is given by $\tau^{-1} = \tau^{-1}_{\rm N} + \tau^{-1}_{\rm B}$ where $\tau_{\rm N} = \tau_0 \cdot \exp(\tfrac{KV}{k_{\rm B}T})$ with a pre-exponential factor $\tau_0 \approx 10^{-13} - 10^{-9}\,$s is the N{\'e}el relaxation time, and $\tau_{\rm B} = \tfrac{3V_{\rm H}\eta}{k_{\rm B}T}$ is the Brown relaxation time arising from the MNPs rotation in the liquid with the viscosity $\eta$ and a hydrodynamic MNP volume $V_{\rm H} > V$.

The heating efficacy of magnetic nanoparticles is typically given by the specific loss parameter SLP in Watts per gramm magnetic material as
\begin{equation}
\rm{SLP} := \frac{P}{\rho} = \frac{Af}{\rho}\,\, ,
\label{eq:SLP}
\end{equation}
where $\rho$ is the density of the used magnetic material.

Note that both relaxation mechanisms contribute to the total power loss but the latter remains directly related to the hysteretic loss area $A$ of the $M(H)$-loop occurring at the frequency $f$ of the AMF\,\cite{Helbig2023}. LRT however overestimates the power loss. For this reason, and also to incorporate inter-particle interactions, Ruta et al.\,\cite{Ruta2015} developed a kinetic Monte Carlo model to calculate the SLP for a suspension of SPIONs with a radii $r=7$ to 20\,nm using a saturation magnetization of 400\,kA/m and an anisotropy $K=30\,$\,kJ/m$^3$ typical for magnetite. A power loss of about 300\,W/g was obtained for an AMF frequency $f=100\,$kHz and a field amplitude $\mu_0 H = 30\,$mT. Note that for non-interacting MNPs SLP values up to 670\,W/g (the SLP value given in Fig.\,\ref{Fig:limitations}\textbf{a}) have been obtained for an optimized particle diameter $D=13.8\,$nm. It is noteworthy that an ensemble of MNP having a finite size distribution will exhibit a markedly reduced SLP value. This reduction is primarily because the majority of MNPs possess diameters that deviate from the optimum size required for a specific frequency and field intensity of the alternating magnetic field (AMF). 

It is further noteworthy that the product of the field amplitude $H$ and frequency $f$ remains limited because in a hyperthermia application, healthy body tissue not loaded with the magnetic particles may overheat due to Eddy current losses. Hergt and Dutz\,\cite{Hergt2007} have reported a biological discomfort level (BDL) as $H\cdot f< 5\cdot 10^9$Am$^{-1}$s$^{-1}$. Slightly lower limits of 1.8 or 2 $\cdot 10^9$Am$^{-1}$s$^{-1}$, have been reported by Jordan et al.\,\cite{Thiesen2008} for the treatment of prostate tumor patients and Mamiya et al.\,\cite{Mamiya2013} considering the cooling by the blood flow. Here we will use the BDL by Hergt and Dutz\,\cite{Hergt2007}. We strongly emphasize that such a BDL should be considered when comparing the heating efficacy of different SPNPs, which is unfortunately often ignored or not even stated in many previous works. With this in mind one may for example consult the table of the SLP values given in Gavil{\' a}n et al.\,\cite{Gavilan2021}, which, considering this limit, is in good agreement with the maximum SLP estimated by Ruta et al.\,\cite{Ruta2015} for SPIONs. 

Compared to chemical synthesis, top-down fabrication of particles offers a large variety of design options in view of the choice of materials, magnetic properties, but also concerning the particle size and shape which can be defined by typical micro-fabrication approaches. Layers of magnetic materials, along with additional layers, can be employed for various purposes, such as achieving optimized growth conditions, enhancing interfacial anisotropies, or facilitating magnetic interlayer coupling. Additionally, sacrificial layers may be utilized to enable the subsequent separation of disk-shaped islands. These layers and architectures thereof play a critical role in tailoring the physical and magnetic properties of the structures for specific applications. In addition, further layers may be introduced for oxidation protection or for facilitating a successive (bio)chemical functionalization. To achieve stable suspensions of such disk-shaped particles, the total magnetic moment at zero field must vanish. This can be accomplished with synthetic antiferromagnet disk particles (SAF-MDP) consisting of two ferromagnetic layers (FL) that are antiferromagnetically (AF) coupled. 
The AF-coupling between the two FLs can be obtained either through the stray fields of the two F layers\,\cite{Hu2008} for the case of SAF-MDP with in-plane magnetization or by an antiferromagnetic Ruderman–Kittel–Kasuya–Yosida (RKKY) interaction\,\cite{Ruderman1954,Kasuya1956,Yosida1957}, for example occurring by a 0.6\,nm-thick Ru interlayer, as typically employed for SAF-MDP with perpendicular magnetization. 

However, the $M(H)$-loops of all these SAF-MDPs with an in-plane magnetization showed an almost linear $M(H)$-loop without any hysteresis reminiscent to a Stoner-Wohlfarth\,\cite{Stoner1948} like hard axis magnetization process. Hu et al.\,\cite{Hu2008,Hu2009} attributed this to a spin-flop process, where the initially antiparallel magnetization directions of the two FLs flip away from $\textbf{e}_{K_{\rm u}}$ for fields $H\hspace{-0.7mm}\parallel\hspace{-0.5mm} \textbf{e}_{K_{\rm u}} > H_{\rm sf}$, where $H_{\rm sf}$ is the spin-flop field, and gradually scissor towards $\textbf{e}_{K_{\rm u}}$ for increasing fields, leading to an almost linear and hysteresis-free $M(H)$-loop. Hence SAF-MDP with in-plane have so far only been designed for applications like separation of biomolecules or cell manipulation and sorting\,\cite{Zhang2013} but because of the lack of a magnetic hysteresis are not suited for hyperthermia applications. 

Vemulkar et al.\,\cite{Vemulkar2015} fabricated SAF-MDPs with perpendicular magnetization states. In liquid, these particles align their hard axis to the field for fields up to about 140 to 300 mT, resulting in a linear M(H) loop without hysteresis. At higher fields, the SAF-MDPs align their easy axis to the applied field, accompanied by a small hysteretic loss area. Consequently, the total hysteretic losses remain small, rendering these particles unsuitable for hyperthermia applications. Subsequent research\,\cite{Vemulkar2017} focused on the mechanical response of such perpendicular SAF-MDPs in liquid for magnetomechanical cancer cell destruction\,\cite{Cheng2016,Mansell2017}.
However, 
disk-shaped nanoparticles comprised of only one FL (using inexpensive magnetic materials) forming a magnetic vortex state have also been used for magnetomechanical cancer cell destruction\,\cite{Kim2010,Leulmi2015,GoirienaGoikoetxea2020,Naud2020}. 
SAF-MDPs with larger 
easy axis hysteretic losses have recently been reported by Li et al.\,\cite{Li2022} for disk diameters of 1.8\,$\mu$m and 120\,nm. However, the switching fields were too high for Hyperthermia applications and the shape of the $M(H)$ loop is rounded by wide switching field distributions typically occurring patterned films with perpendicular magnetization\,\cite{Adhikari2023,Jamet1998,Thomson2006,Shaw2007,Lau2008,Pfau2011}. 

In our work, we establish the fundamental limits for the maximum achievable hysteretic losses and employ micromagnetic and fluid-mechanical modeling together with single particle micromagnetic characterization to obtain SAF-MDPs with maximized hysteretic losses opening an avenue to reach the absolute limits of the heating power given by physics. 

\section*{Results and Discussion}
\subsection*{Maximizing Hysteretic Losses for in-plane SAF-MDPs}
\begin{figure}[t]
\centering
\includegraphics[width = 85mm]{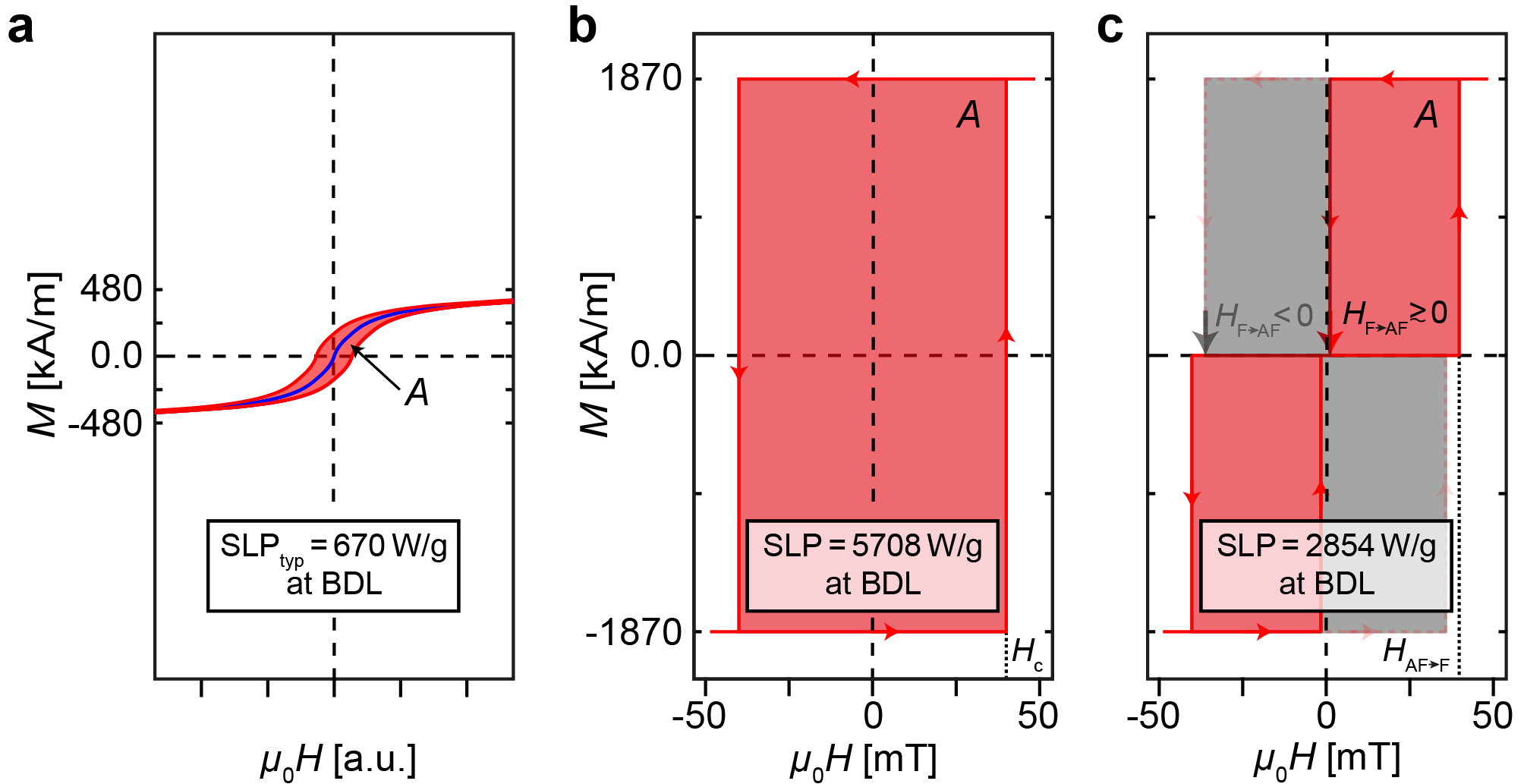}
\caption{\textbf{Physics-based limits of specific loss parameters under the boundary conditions of the biological discomfort level. a} $M(H)$-loop of a superparamagnetic nanoparticle particle showing Langevin-like loop without hysteresis for slowly varying fields (blue curve) and developing a small hysteretic loss (red-shaded area) for fields oscillating at higher frequencies. \textbf{b} ideal $M(H)$-loop of a ferromagnetic disk particle (F-MDP) leading to the largest hysteretic loss (red-shaded area) of 5708\,W/g$_{\rm CoFe}$ possible under consideration of the biological discomfort level. \textbf{c} ideal $M(H)$-loop of a synthetic antiferromagnet (AF) disk particle (SAF-MDP) with $m(H=0) = 0$ (red-shaded area). The maximum SLP is half the size of that of the F-MDP, but can be increased for a SAF-MDP with a reduced antiferromagnetic (AF) coupling. Then the magnetic moment does not longer vanish at zero field, and a demagnetization procedure has to be applied to re-set the SAF-MDP into its af-coupled ground state. }
\label{Fig:limitations}
\end{figure}

The hysteretic loss of any system is fundamentally limited by the material with the highest saturation magnetization in Nature, which for the case of magnetic CoFe thin films is $M_{\rm s} = 1870\,$kA/m\,\cite{Liu2008} . For a perfectly rectangular hysteresis loop (Fig.\,\ref{Fig:limitations}\textbf{b}) a ${\rm SLP}_{\rm F,max} = 4\mu_0\cdot M_{\rm s}  (H_{\rm c}\cdot f) / \rho_{\rm CoFe} = 5708\,$W/g$_{\rm CoFe}$ would then be obtained.
Note that this is the highest SLP possible by the limits of Nature under the boundary condition of the BDL as given by Hergt and Dutz\,\cite{Hergt2007} and that for a system reaching its saturation magnetization, the SLP is independent of the size of the coercive field. This is because within the boundaries of the BDL a larger $H_{\rm c}$ leads to a correspondingly smaller $f$. 

The maximum SLP for superparamagnetic magnetite particle (SPIONs) with $M_{\rm s} = 480\,$kA/m must hence be much smaller than 1747\,W/g$_{{\rm Fe}_3{\rm O}_4}$ or 2413\,W/g$_{\rm Fe}$ which would be obtained only for a perfectly rectangular $M(H)$-loop. 
However, because magnetic nanoparticles with a finite anisotropy are only superparamagnetic if the energy barrier between the two magnetic states can be overcome by the thermal energy available at room temperature, the switching field distribution of an ensemble of superpamagnetic particles for an applied AMF is inherently wide even for mono-disperse particles, and thus always yields a non-rectangular hysteresis loop\,\cite{suess2016superior}.
Additionally, due to thermal activation, saturation magnetization cannot be achieved in AC field cycles. For example, Co-like superparamagnets with diameters of 13\,nm, subjected to an AC field with amplitude of 5\,mT and a frequency of 100\,kHz, only reach a magnetization of about 50\% of the saturation magnetization $M_s$\,\cite{Ruta2015}. As shown in\,\cite{Ruta2015} the obtained hysteresis loops are elliptical like, which can not utilize the entire area of the rectangle with area $4\mu_0 H_c M_s$. 

We thus argue that an anisotropy energy substantially exceeding thermal energy and thus a sufficiently large particle volume is a necessary but not yet sufficient condition to obtain the desired rectangular hysteresis loop. For suspension stability, the particle must further have a vanishing total magnetic moment at zero field. Both conditions can in principle be met by a synthetic antiferromagnet disk particle (SAF-MDP) consisting of two antiferromagnetically (AF) coupled CoFe FLs with equal magnetic moments exhibiting a hypothetical hysteresis loop as depicted in Fig.\,\ref{Fig:limitations}\textbf{c}.
For fields $H\geq 0$, the ideal $M(H)$-loop for such a SAF-MDP is characterized by a field $H>H_{\rm AF\rightarrow F}$, where the two FLs switch from the AF-coupled ground state into a ferromagnetically (F) aligned state, and a field $H_{\rm F\rightarrow AF} \gtrsim  0$ below which the SAF-MDP returns back into its AF-coupled ground state. 

The maximum absorption loss rate of an ideal SAF-MDP then is equal to $2854\,$W/g$_{\rm CoFe}$, half of that of a ferromagnetic disk particle. Note however, that $m_{\rm SAF-MDP}(H=0)=0$ is not a necessary condition for a stable suspension, because SAF-MDPs with a $H_{\rm F\rightarrow AF}<0$ may have still an AF-coupled ground state with a vanishing magnetic moment, which can be reached with a suitable demagnetization procedure.

However, to enable a suspension of SAF-MDP with in-plane magnetization to exhibit hysteretic loss, two primary issues must be addressed: preventing the spin-flop process and developing a method to align the particles' easy magnetization axis with the applied AMF. Addressing the first issue involves designing FLs with high in-plane uniaxial anisotropy.

\subsection*{Micromagnetic Modeling}
\begin{figure}[t]
\centering
\includegraphics[width = 85mm]{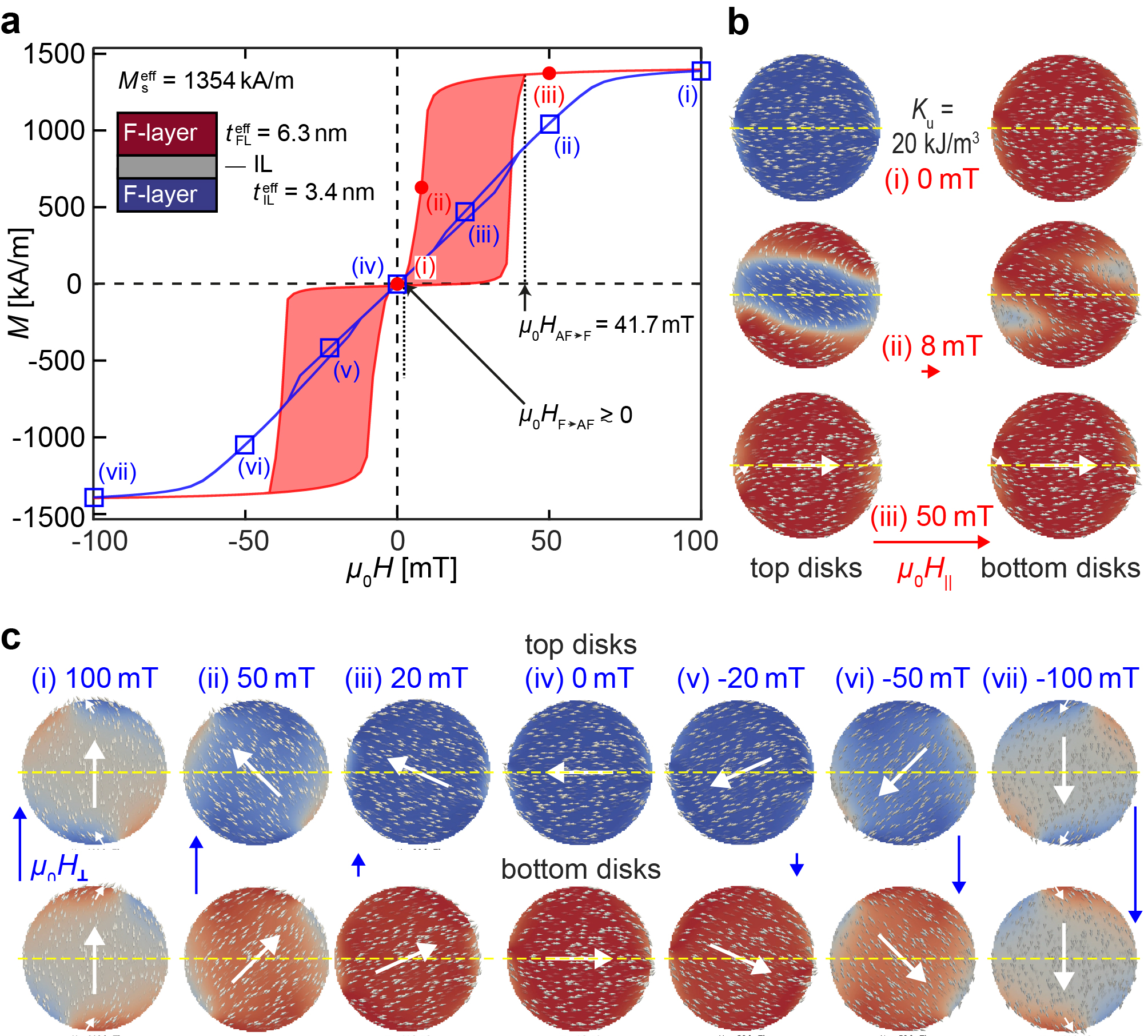}
\caption{\textbf{Micromagnetic modeling based design of an in-plane SAF-MDP with 20\,kJ/m$^3$ uniaxial anisotropy a} easy (red curve) and hard axis (blue curve) $M(H)$-loop with large and vanishing hysteretic loss. The top left inset displays the SAF layer structure. \textbf{b} micromagnetic states of the upper and lower F layer for fields of 0\,mT, 8\,mT, and 50\,mT applied along the easy axis (dashed yellow lines). \textbf{c} micromagnetic states of the upper and lower F layer for fields of 100\,mT, 50\,mT, 20\,mT, 0\,mT, -20\,mT, -50\,mT and -100\,mT  revealing the hysteresis-free rotation of the magnetic moments for fields applied perpendicular to the easy axis (yellow lines).}
\label{Fig:micromagmodel}
\end{figure}
The micromagnetic state of small disk-shaped magnetic particles is governed by various parameters, including the disk diameter, thickness, and the intrinsic properties of the magnetic layer such as magnetization, exchange stiffness, and magnetic anisotropy. Typically, small disks with a thin, single FL form a uniform mono-domain state, as documented by Cowburn et al.\,\cite{Cowburn1999}. In contrast, larger disks tend to exhibit multi-domain or vortex states, as noted by Usov et al.\,\cite{Usov1993} and Guslienko et al.\,\cite{Guslienko2001}. Disks with intermediate sizes are expected to maintain a mono-domain state, albeit with magnetic moment directions locally deviating from the easy in-plane axis.
To achieve a hysteretic easy axis magnetization process characterized by a small field 
$H_{\rm AF\rightarrow F}$ compatible with the alternating magnetic field (AMF) amplitudes typically encountered in hyperthermia experiments, along with a narrow switching field distribution and a field 
$H_{\rm AF\rightarrow F} > 0$, we carried out comprehensive micromagnetic modeling. Our aim was to identify a magnetic parameter space that would establish an $M(H)$-loop for SAF-MDPs with 500\,nm diameter demonstrating stability against minor fluctuations in magnetic properties. This endeavor was geared towards ensuring the reliability and consistency of the system's performance. 

Fig.\,\ref{Fig:micromagmodel}\textbf{a} illustrates the hysteretic easy axis $M(H)$-loop (red curve) together with the (essentially) non-hysteretic hard axis $M(H)$-loop (blue curve) obtained from our micromagnetic modelling for an SAF-MDP with a diameter of 500\,nm, an effective FL thickness of $t^{\rm eff}_{\rm FL} = 6.3\,$nm, an effective interlayer thickness of $t^{\rm eff}_{\rm IL}=3.4\,$nm, a uniaxial in-plane anisotropy of 20\,kJ/m$^3$, and an effective saturation magnetization $M_{\rm s}^{\rm eff} = 1354\,$kA/m (here the effective thicknesses and saturation magnetization describes the quantities as found for our experimental SAF-MDP system). 
Fig.\,\ref{Fig:micromagmodel}\textbf{b} displays the micromagnetic states for fields of 0\,mT, 8\,mT and 50\,mT, demonstrating that the substantial uniaxial anisotropy present here (along the dashed yellow lines) can effectively suppress the spin-flop process, thereby enabling a hysteretic easy axis magnetization loop. Note also that an anisotropy much higher than 20\,kJ/m$^3$ is unfavorable, as it would increase the $H_{\rm AF\rightarrow F} = 41.7\,$mT switching field and thus the required AMF amplitude, and prevent the SAF-MDP from returning to its AF ground state at zero field. Additionally, it would lead to a reduction of the domain wall thickness from the present $\delta_{\rm dw} = \pi\sqrt{\tfrac{A}{K_{\rm u}}} = 82\,$nm, where $A\approx 15\,$pJ/m is the exchange stiffness assumed for a typical FL material. For a real SAF-MDP with defects, this could lead to undesirable multi-domain states and increased domain wall pinning, broadening the transition between AF and F-states as for example observed for perpendicular SAF-MDP\,\cite{Li2022}. Fig.\,\ref{Fig:micromagmodel}\textbf{c} then illustrates the rotatory, (essentially) hysteresis-free $M(H)$-loop (blue curve) obtained for fields applied perpendicular to the easy axis (dashed yellow lines).

\subsection*{Materials Selection and MDP Fabrication}\label{sec3}
\begin{figure*}
\centering
\includegraphics[width = 175mm]{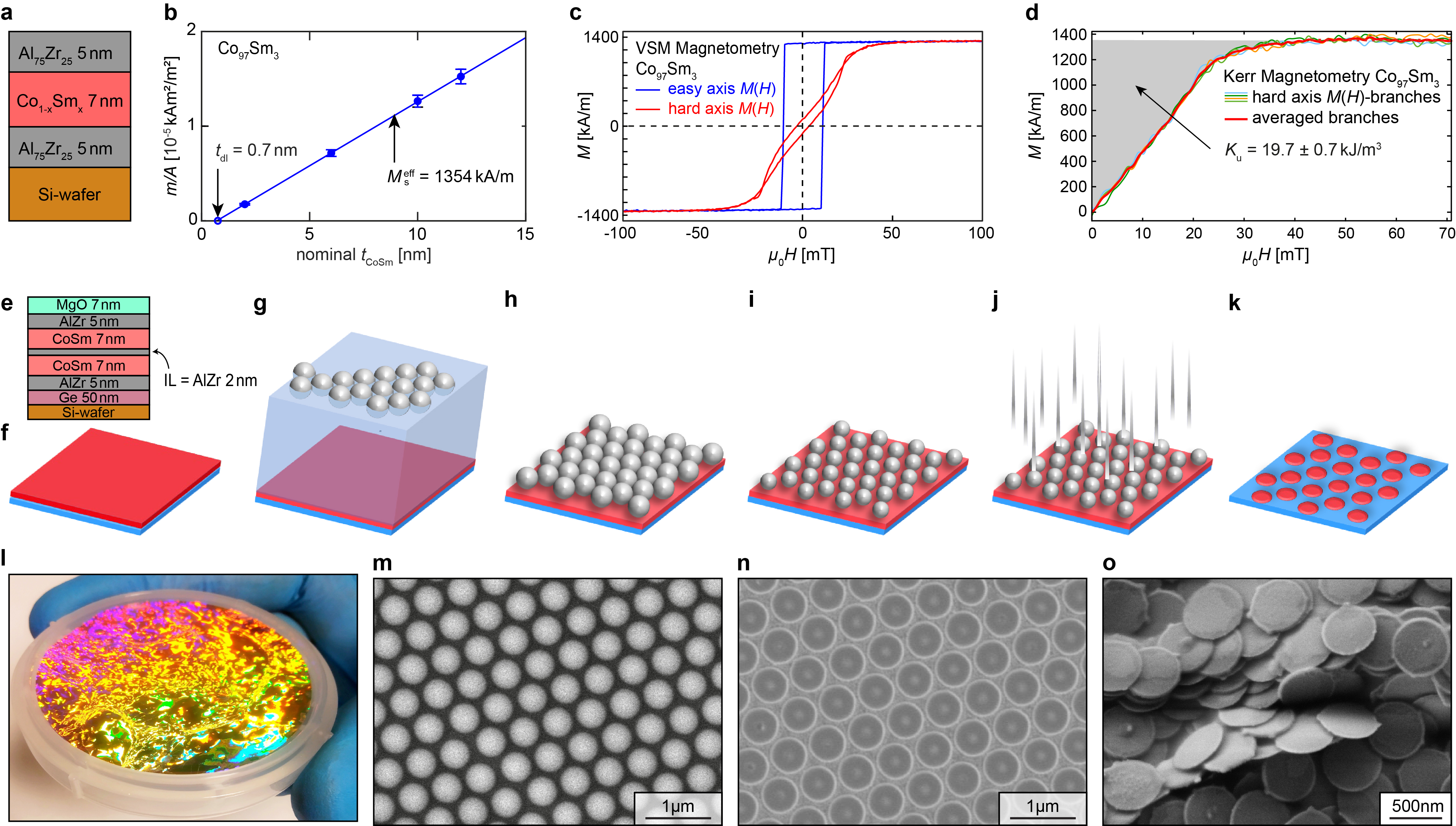}
\caption{\textbf{Materials selection and SAF-MDP fabrication. a} multilayer system containing a single Co$_{1-x}$Sm$_x$ layer as used for the determination of the dead layer thickness, and the evaluation of the anisotropy. \textbf{b} magnetic moment of Co$_{97}$Sm$_{3}$ layers as a function of the layer thickness as used for the determination of the dead layer thickness, and saturation magnetization. \textbf{c} easy (blue curve) and hard (red) axis VSM data. \textbf{d} Kerr magnetometry results used for the determination of the uniaxial anisotropy. \textbf{e} schematic of the SAF multilayer structure. \textbf{f} to \textbf{k} schematics of the nanopatterning process employed in our work to fabricate the SAF-MDP islands still attached to the wafer substrate \textbf{k}. \textbf{l} wafer with the SAF-MDP multilayer structure and self-assembled polystyrene (PS) sphere as schematically depicted in \textbf{h}. \textbf{m} scanning electron microscopy (SEM) image of the self-assembled PS spheres with a diameter of 500\,nm reduced from the initial diameter of 660\,nm by oxygen plasma etching. \textbf{n} and \textbf{o} SEM pictures of the SAF-MDP attached to the wafer and re-deposited onto a substrate from an SAF-MDP suspension, respectively. }
\label{Fig:fabrication}
\end{figure*}
While SAF-MDP with large perpendicular anisotropies can be obtained by interfacial anisotropies occurring for example in Co/Pt multilayers\,\cite{Vemulkar2015},  achieving a large uniaxial anisotropy in the order of 20\,kJ/m$^3$ in in-plane magnetized systems is more challenging. An in-plane uniaxial anisotropy of about 25\,kJ/m$^3$ could for example be achieved for a 6\,nm-thick CoFeB layer sputter-deposited onto a 15\,nm-thick Ta seed layer fabricated by oblique sputter-deposition with an angle of 60$^{\circ}$\,\cite{Scheibler2023}. However, the obtained anisotropy is of interface type and hence decays with the inverse thickness of the FL making it challenging to obtain an equal anisotropy in both FLs of an SAF-MDP. Here, we thus perform an in-field sputter-deposition of amorphous Co$_{1-x}$Sm$_x$ layers for which anisotropies up to about 200\,kJ/m$^3$ were obtained for $x=0.2$\,\cite{Magnus2013}. For the 20\,kJ/m$^3$ required according to our micromagnetic modeling work, much smaller concentrations of Sm were tested and an optimal concentration of 3\% was determined. Further, to promote the amorphous growth, Al$_{70}$Zr$_{30}$ layers were used for the 5\,nm-thick seed layer, 2\,nm thick interlayer and for the 5\,nm-thick oxidation protection layer. 
To obtain the total thickness of magnetic dead layers forming on the oxide,  we fitted the magnetic moment as a function of increasing thickness of a Co$_{97}$Sm$_3$ films with a straight line  (Fig.\,\ref{Fig:fabrication}\textbf{b}). Its interception with the thickness axis then corresponds to the thickness of the total magnetic dead layer ($t_{\rm dl} = 0.7\,$nm) forming at the interfaces to the adjacent Al$_{70}$Zr$_{30}$ layers resulting in an effective thickness of the FL hence $t^{\rm eff}_{\rm FL} = 7\,{\rm nm} - t_{\rm dl} = 6.3$\, nm, while the slope of 1354\,kA/m corresponds to the effective saturation magnetization $M^{\rm eff}_{\rm s}$. The dead layer thickness also enhances the interlayer thickness from the nominally 2\,nm to 3.4\,nm. 
$M(H)$-loops for a nominally 7\,nm-thick Co$_{97}$Sm$_3$ film were acquired by VSM  with the easy (blue curve) and hard (red) axis loops diplayed in  Fig.\,\ref{Fig:fabrication}\textbf{c}. The hard axis $M(H)$-loop deviates from the expected ideal linear and hysteresis free $M(H)$-loop which can be attributed to hysteretic processes occurring near the sample edges. A perfectly linear and hysteresis free hard axis $M(H)$-loop could however be obtaind by  Kerr magnetometery (Fig.\,\ref{Fig:fabrication}\textbf{d})
which permits localized measurements in the center of a larger sample with a fine adjustment of the angle of the applied field but cannot obtain a value for $M_{\rm s}$. Using the $M_{\rm s}^{\rm eff}$ from the VSM measurement, an anisotropy $K_{\rm u}= (19.7\pm 0.7)\,$kJ/m$^3$ close to the 20\,kJ/m$^3$ used for the micromagnetic modeling (Fig.\,\ref{Fig:micromagmodel}\textbf{a}) was determined from the area enclosed between the magnetization axis and the averaged four 0-to-$M_{\rm s}$-branches of the $M(H)$ Kerr magnetometry data.

For the fabrication of the SAF-MDP, the multilayer structure depicted in Fig.\,\ref{Fig:fabrication}\textbf{e} was sputter-deposited onto a 50\,nm thick Ge sacrificial layer. The Ge layer was previously deposited onto the wafer substrate by thermal evaporation in a separate chamber. The top-most 7\,nm-thick MgO layer serves as an independent sacrifical layer later used to remove remainders arising from the patterning process. 

For the patterning a polystyrene-based lithography approach adapted
from Giersig et al.\,\cite{Rybczynski2003} was employed with additional process improvements introduced by other authors \cite{ArmstrongEileen2015, Krupinski2019, Qiu2022} (Fig.\,\ref{Fig:fabrication}\textbf{f} to \textbf{k}). This approach offers high scalability to the size of several in$^2$ and ensures good repeatability at minimal cost since no special equipment is required. For this, polystyrene (PS) beads prepared via free radical initiated polymerization with a diameter of 660\,nm were self-assembled at the water-air interface, creating a highly ordered hexagonally packed monolayer of PS submicron spheres (Fig.\,3\textbf{l})\,\cite{Krupinski2019,GoirienaGoikoetxea2016,Welbourne2021-1}. The quality of the self-assembled periodic pattern of PS beads over a full 2-inch Si wafer was confirmed through visual inspection (Fig.\,3\textbf{l}). Following oxygen plasma reactive ion etching to separate the PS spheres and reach a bead diameter of 500\,nm (Fig.\,3\textbf{i} and \textbf{m}), Ar milling was performed to pattern the wafer (Fig.\,3\textbf{j}). Afterwards the PS beads were removed by ultrasound in water, yielding circular disk-shaped pillars still attached to the wafer (Fig.\,3\textbf{k} and \textbf{n}). The height of the MDPs
on the wafer was determined as approximately 50\,nm and is
higher than the total thickness of the magnetic multilayer (33\,nm). This is
because the ion-milling process was continued into the Ge sacrificial layer
to ensure that the bottom AlZr layer has also been removed. Finally, the
MgO top sacrificial layer was removed using citric acid and the Ge bottom
sacrificial layer was dissolved in a final step to release the SAF MDPs from
the wafer into suspension. Fig.\,3\textbf{o} then shows the particles re-deposited from the suspension onto a wafer surface for successive scanning electron microscopy observation.

\subsection*{Magnetic and Hyperthermia Characterization and extended Micromagnetic and Mechanofluidic Modeling}
\begin{figure*}
\centering
\includegraphics[width = 175mm]{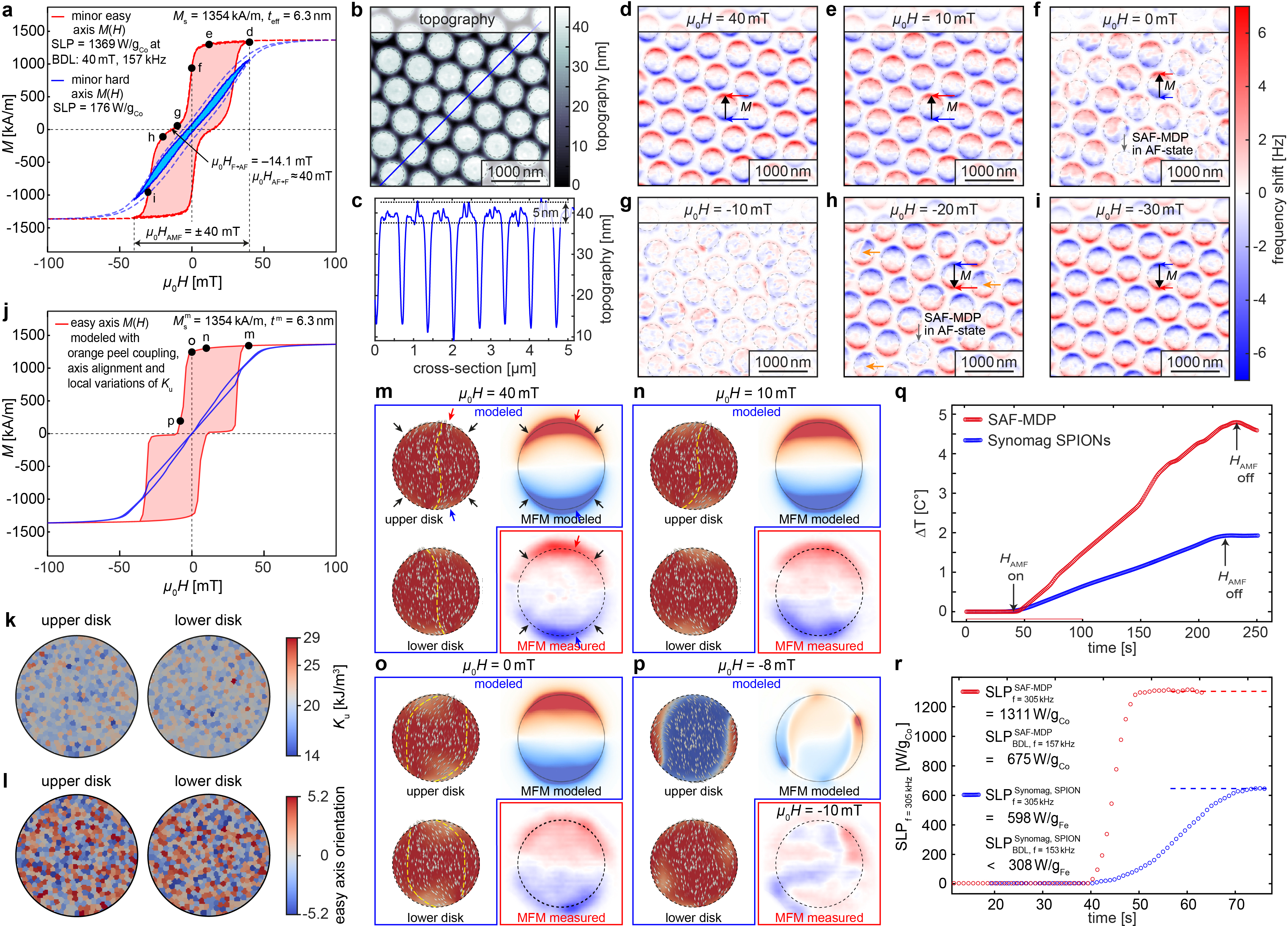}
\caption{\textbf{SAF-MDP magnetic and hyperthermia characterization. a} VSM measurements of the easy axis $M(H)$-loop (solid and dashed red curves, for the minor and major loop, respectively), hard axis $M(H)$-loop (solid and dashed blue curves, for the minor and major loop, respectively) of our SAF-MDPs still attached to the wafer ($D = 500\,$nm, 7\,nm-thick Co$_{97}$Sm$_3$ F layers separated by 2\,nm of Al$_{70}$Zr$_{30}$). The labels \textbf{\small d} to \textbf{\small i} refer to the corresponding MFM data displayed in panels \textbf{d} to \textbf{i}. \textbf{b} topography image and \textbf{c} cross-section of the SAF-MDP islands. \textbf{d} to \textbf{i} magnetization reversal of the SAF-MDP studied by magnetic force microscopy (MFM) performed in fields applied along the easy axis, ranging from $40\,$mT to $-30\,$mT. \textbf{j} easy and hard axis (red and blue lines, respectively) $M(H)$-loops obtained from a refined micromagnetic model using distinct distributions anisotropies and easy axes alignments of the grains in the upper and lower disks as displayed in \textbf{k} and \textbf{l}, respectively, and employing a roughness-induced ferromagnetic orange-peel coupling between the FLs. Labels \textbf{m} to \textbf{p} refer to the modeling results compared to experimental results displayed in panels \textbf{m} to \textbf{p}. 
\textbf{m}, \textbf{n}, \textbf{o}, and \textbf{p} micromagnetic states of the top and bottom disks (right side) together with modeled (top right) and measured (bottom right) MFM data for fields ranging from 40, 10, 0 and $-10$\,mT. \textbf{q} temperature rise as a function of time for suspensions of 80\,$\mu$g$_{\mathrm{Co, Fe}}$/ml of SAF-MDP and Synomag superparamagnetic nanoparticles. \textbf{r} evolution of the SLP values with time for suspensions of the AF-MDP and Synomag superparamagnetic nanoparticles calculated from \textbf{q}.}
\label{Fig:results}
\end{figure*}

Following the fabrication, the obtained SAF-MDP were characterized. Fig.\,\ref{Fig:results}\textbf{a} shows the easy axis $M(H)$-loop (solid red and dashed red curve for the minor and major loop, respectively), together with the hard axis $M(H)$-loop (solid blue and dashed blue curve for the minor and major loop, respectively) for the SAF-MDPs still attached to the wafer as measured by VSM. The easy axis field required to drive the SAF-MDP aligned into F-state is about 40\,mT and thus much smaller than the about 70.8\,mT required for the modeled loop (Fig.\,\ref{Fig:micromagmodel}\textbf{a}, red curve) which does however not include temperature effects. Further, compared to the loop modeled for SAF-MDPs (Fig.\,\ref{Fig:micromagmodel}\textbf{a}, red curve and area), the $M(H)$-loop measured by VSM (Fig.\,\ref{Fig:results}\textbf{a}, red curve and area) shows less abrupt transitions between the AF- and F-states indicating the presence of complex domain states during the switching process. 

The hysteretic losses calculated from the minor easy and hard axis (solid red curve / red area and solid blue curve / blue area in Fig.\,\ref{Fig:results}\textbf{a}, respectively.) $M(H)$-loops for a $\pm B=40\,$mT field excursion 
are 77.6 and 9.97\,kJ/m$^3$, respectively. Using eq.\,\ref{eq:SLP}, SLP values of 1369 and 176\,W/g$_{\rm Co}$ are obtained at the BDL for $B=40\,$mT, $f=157\,$ kHz. 

Hence, the easy axis SLP expected from the $M(H)$-loop is only about 48.0\% of the $2854\,$W/g$_{\rm Co}$ calculated for an ideal SAF-MDP (Fig.\,\ref{Fig:limitations}\textbf{c}). 
A reduction of the SLP by 72.4\,\% can be expected from the  
saturation magnetization of 1354\,kA/m of the CoSm ferromagnetic layer used here, which is considerably smaller than the $1870\,$kA/m of CoFe alloy films\,\cite{Liu2008}. Hence, to obtain the observed total SLP reduction of 48.0\%, the remaining 66.2\,\% must be attributed to the more gradual switching from the AF to the F-state. The latter demonstrates the importance of sharp switching between the AF and F-states. 

To gain insight in the nature of the switching process, we employed a home-built magnetic force microscope working under vacuum conditions to obtain increased measurement sensitivity\,\cite{Meyer2021,Feng2022}. 
Advanced techniques to disentangle contrast contributions from the topography (Fig.\,\ref{Fig:results}{\bf b}, cross-section {\bf c}) from those arising from the magnetic field and deconvolution techniques\,\cite{Feng2022,Meyer2021} were employed to obtain the frequency shift data depicted in Fig.\,\ref{Fig:results}{\bf d} to {\bf i} for different fields applied along the in-plane easy axis of the SAF-MDPs. Figure panels\,\ref{Fig:results}{\bf d} and {\bf e} show the SAF-MDPs micromagnetic state at saturation at $\mu_0 H = 40\,$mT and at $\mu_0 H = 10\,$mT. Half-moon shaped red and blue features are apparent at the upper and lower edges of the SAF-MDPs (red and blue arrows in Figs.\,\ref{Fig:results}{\bf d} and {\bf e}). The magnetic force microscopy (MFM) results obtained at zero field (Fig.\,\ref{Fig:results}{\bf f}) show that some of the SAF-MDPs have switched back into their AF ground state but several disks still show a substantial red/blue MFM contrast, indicating an incomplete switching process. The latter is also apparent from the magnetic remanence (point f in Fig.\,\ref{Fig:results}{\bf a}). At a field $\mu_0 H = -10\,$mT the magnetization nearly vanishes (point g in Fig.\,\ref{Fig:results}{\bf a}). The MFM image (Fig.\,\ref{Fig:results}{\bf g}) consequently shows on a very small granular contrast indicating the magnetic moments in the upper and lower disks are essentially antiparallel. Hence the stray fields generated by the opposite poles of the two disks essentially compensate each other. At $-20\,$mT many disks have already switched into their F-states with a south pole (blue half moon) at the top and a north pole (red half moon) at the bottom (Fig.\,\ref{Fig:results}{\bf h}), whereas at $-30\,$mT all disks visible in the MFM image (Fig.\,\ref{Fig:results}{\bf i}) are in the F state, compatible with the $M(H)$-loop which is almost saturated at point i. 

To improve our understanding of the differences between the experimental (Fig.\,\ref{Fig:results}{\bf a}) and initially modeled $M(H)$-loops (Fig.\,\ref{Fig:micromagmodel}{\bf a}), further, more realistic, micromagnetic modeling was performed. For this, the disks were were patterned into about 20\,nm-diameter grains having a Gaussian standard deviation of $\pm 2\,$kJ/m$^3$ around an average anisotropy $K_{\rm u} = 20\,$kJ/m$^3$, and a uniform variation of the anisotropy axes directions by $\pm 3^{\circ}$ (Fig.\,\ref{Fig:results}{\bf k} and {\bf l}). Such local variations of the anisotropy magnitude and easy axis direction lead to a moderate domain wall pinning and explain the observed small particle-to-particle variation of the magnetization switching process
(Figs.\,\ref{Fig:results}{\bf d} to {\bf i}). However, the experimentally observed shift of the field, at which the SAF-MDPs return to their AF ground states after positive saturation to negative field values, is not due to granular variations in anisotropy. Instead, it must result from a reduction in the stray field-generated AF coupling between the two FLs. Such a reduction can for example occur by a ferromagnetic orange-peel coupling arsing from an interlayer with a finite roughness\,\cite{neel1962new}. Here a ferromagnetic orange-peel exchange $J_{\rm op} = -t^{\rm eff}_{\rm FL}\mu_0 M^{\rm eff}_{s} H_{\rm shift} = 0.064\,$mJ/m$^2$ has been used in our refined micromagnetic model calculation to obtain a good match between the modeled (Fig.\,\ref{Fig:results}{\bf j}) and experimentally (Fig.\,\ref{Fig:results}{\bf a}) observed easy axis $M(H)$-loops. Note that the field $\mu_0 H_{\rm shift} \sim -15\,$mT is the difference of the shift of the centers of the $M(H)$-loops with and without orange-peel coupling towards negative fields. 

Fig.\,\ref{Fig:results}{\bf m} to {\bf p} then display the modeled magnetic moment distributions of the top and bottom disks (left side, top and bottom images) together with a comparison of the simulated and measured MFM images (right top and bottom images) for different applied fields. For this, specific SAF-MDP were selected for each field to achieve a reasonable match with the modeled MFM data. We note that this is justified as the real defect distribution in a specific SAF-MDP island remains unknown, and hence only a qualitative agreement can be expected. For a field of 40\,mT both FLs are almost saturated, and the magnetic moments are well aligned to the direction of the anisotropy axis direction (yellow dashed curves) with a slight s-shaped orientational deviation from the easy axis. The red and blue half moons (Fig.\,\ref{Fig:results}{\bf m}, MFM model and measurement) arise from the positive and negative magnetic charges generated by the divergence of the magnetization field at the upper and lower disk boundaries, respectively. Note that the divergence is strongest if the local orientation of the magnetic moment vector is perpendicular to the disk boundary (at the top and bottom centers, red and blue arrows) and weaker at the disk sides (black arrows). When the applied field is reduced, the stray field of one disk acting on the moments of the other disk leads to a gradual deviation of the magnetic moments away from the easy magnetization axis at 10 and 0\,mT(see increased curvatures of the dashed yellow lines in Fig.\,\ref{Fig:results}{\bf n} and {\bf o}, upper and lower disks) and finally to the formation of reversal domains at -8\,mT (Fig.\,\ref{Fig:results}{\bf p}) and consequently to a low moment magnetic state close to the ideal AF-oriented ground state. The dotty features visible for some of the SAF-MDPs (Fig.\,\ref{Fig:results}{\bf n} arise from local divergences of the magnetization field. Compatible with the experimental $M(H)$-loop, at $-10\,$mT, the SAF-MDP has switched back almost perfectly into its AF-aligned ground state (Fig.\,\ref{Fig:results}{\bf p}), here (modeled at -8\,mT) with the magnetic moments of the upper disk mainly pointing point down (blue color)  apart from a small and narrow domain at the right side, and the lower disk pointing up (red color). The simulated MFM contrast then is very weak agreeing well with that observed in the experiment. 

For the measurement of the specific loss parameter, an LCC resonance circuit with impedance matching  permitting the application of ac-fields (AMF) with an amplitude up to 50\,mT at its resonance frequency of 305\,kHz was developed (see supporting materials for a more detailed description).

For the quantification of the SLP of our SAF-MDPs in vivo-relevant settings, SAF-MDPs were harvested from two 2-inch wafers (Fig.\,\ref{Fig:fabrication}{\bf l}) fully covered with hexagonal patterns of SAF-MDP islands (Fig.\,\ref{Fig:fabrication}{\bf n}). The SAF-MDP islands have been removed from the wafer substrate by dissolution of the Ge sacrificial layer in 35\% H$_2$O$_2$, collected by a permanent magnet, while the aqueous phase was replaced by deionized water four times to finally achieve a suspension of SAF-MDPs. We find a yield per wafer of 134\,$\mu$g of SAF-MDP mass, equivalent to 66\% of the
theoretically possible maximum, assuming a perfect close-packed hexagonal pattern of SAF-MDP islands over the entire wafer and a complete recovery of all particles. For the SLP measurements, a suspension of 1\,ml with 134\,$\mu$g of SAF-MDP consisting of a concentration of 80\,$\mu$g$_{\rm Co}$/ml was used. The temperature of the suspension was adjusted to match the 20$^{\circ}$C of the coolant liquid used for the coil of the SLP apparatus to prevent a parasitic heat heat loss. Then, an AMF of 40\,mT and 305\,kHz was applied. The obtained time dependence of the temperature rise of the SAF-MDP suspension is plotted in Fig.\,\ref{Fig:results}{\bf q} (red circles) together with that obtained for a suspension of Synomag\,\cite{thno86759,SynomagVogel} SPIONs with an Fe concentration of also 80\,$\mu$g/ml (blue circles). Clearly, the temperature rise of the SAF-MDP is noticeably faster than that of the Synomag SPION particles. The SLP as a function of the first 60\,s, i.e. 20\,s beyond the time where the AMF was turned on is plotted in Fig.\,\ref{Fig:results}{\bf r}, again for the SAF-MPD (red circles) and the Synomag SPIONS (blue circles). For the SAF-MDP a SLP of 1311\,W/g$_{\rm Co}$ was obtained while that of the SP was only 598\,W/g$_{\rm Fe}$. However, note that the 40\,mT AMF applied at a frequency of 305\,kHz results in a field-frequency product of $9.7\cdot 10^{9}\,$Am$^{-1}$s$^{-1}$ which is almost twice as large as the BDL. Consequently, at the BDL, for a frequency $f = 157\,$kHz the SAF-MDP and Synomag particles would generate a SLP of about 675\,W/g$_{\rm Co}$ and $< 309\,$W/g$_{\rm Fe}$, respectively. Note that for the SAF-MDP the scaling of the SLP is linear with the frequency, while for the superparamagnetic Synomag particles the area of the $M(H)$-loop shrinks at lower frequencies, leading to a further reduction of the SLP. The 675\,W/g$_{\rm Co}$ obtained for the SAF-MDP at the BDL is about 49.3\,\% of the SLP obtained at the BDL from the area of the easy axis $M(H)$-loop measured by VSM (red area in Fig.\,\ref{Fig:results}{\bf a} which is 1369\,W/g$_{\rm Co}$ but about a factor of 7.8 larger than the SLP obtained from the hard axis $M(H)$ loop measured by VSM (blue area in Fig.\,\ref{Fig:results}{\bf a} which is 176\,W/g$_{\rm Co}$. 
\begin{figure*}
\centering
\includegraphics[width = 175mm]{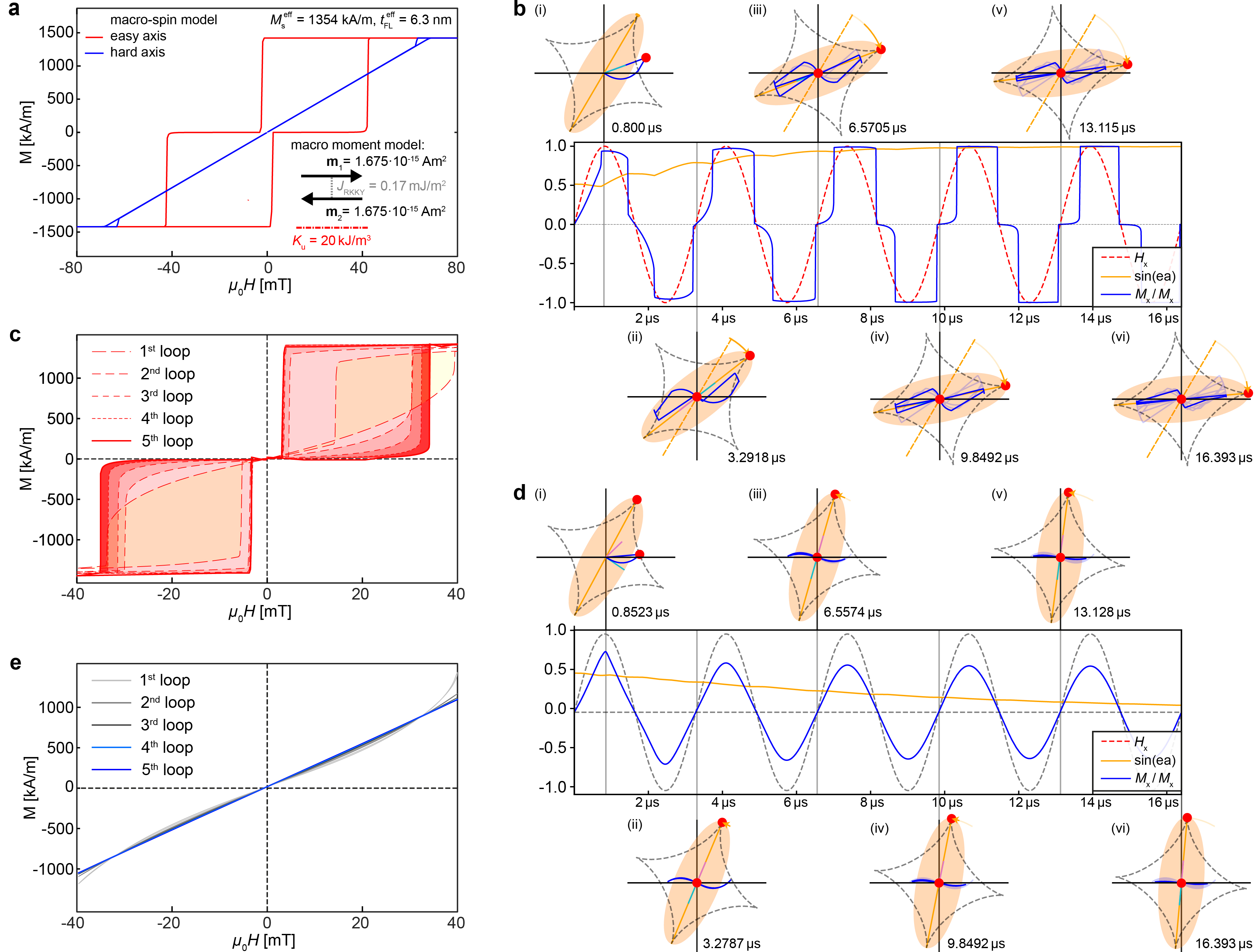}
\caption{{\bf { }|{ }Mechanofluidic modeling of the SAF-MDP alignment to the AMF. a} easy (red curve) and hard axis (blue curve) $M(H)$-loops obtained from a simplified model comprised of two macroscopic magnetic moments ${\bf m}_{1,2}=1.675\cdot10^{-15}\,$Am$^2$ corresponding to the FL magnetic moment, AF coupled with $J_{\rm RKKY}= 0.17\,$mJ/m$^2$ and experiencing a uniaxial anisotropy $K_{\rm u}=20\,$kJ/m$^3$. {\bf b} time evolution of the alignment of the particle's easy axis relative to the axis of the AMF (applied along the horizontal axis) for an initial angle of the easy axis of $\vartheta = 59^{\circ}$. The latter is below the critical angle for which  the particle no longer aligns its easy axis to the AMF. 
The orange ellipses (i) to (vi) showcase the evolution of the particle's easy axis with the time of the applied AMF, while the small purple and turquoise lines inside the ellipses indicate the direction of the two magnetic moments ${\bf m}_{1,2}$ and the evolution of the total magnetic moment, respectively. The diagram between the upper three and lower three ellipses plots the amplitude of the AMF (dashed red curve), the evolution of the easy axis alignment to the AMF (orange curve), and the component of the total magnetic moment along the AMF axis (blue curve). 
{\bf c} evolution of the $M(H)$-loop and hysteresis. {\bf d} time evolution of the alignment of the particle's hard axis parallel to the axis of the AMF for an initial angle $\vartheta = 60^{\circ}$ above the critical angle. {\bf e} decay of the hysteresis of the $M(H)$-loop with time as the particle aligns its hard axis to the AMF. }
\label{Fig:microfluidics_ac_align}
\end{figure*}

We note that the determined SLP is compatible with that obtained if about 41.8\% of the SAF-MDP aligned their easy-axis and the rest of the particles aligned the hard-axis with the direction of the applied AMF. This differentiates our approach from earlier work, where either no\,\cite{Hu2008,Hu2009} or only very weak\,\cite{Vemulkar2015, Vemulkar2017} hysteretic losses were observed. For the case of the SAF-MDP with in-plane magnetic moments and without a well-define uniaxial anisotropy, the absence of the hysteretic loss was attributed to the occurrence of a spin-flop of the AF-coupled magnetic moments, followed by a hard-axis magnetization process\,\cite{Hu2008,Hu2009}. For the SAF-MDP with a strong perpendicular anisotropy\,\cite{Vemulkar2015,Vemulkar2017}, the SAF-MDP tended to align their hard axis with the AMF. 

To elucidate the significant hysteretic losses observed in our experiments, which we attributed to an at least partial alignment of the easy axes of the SAF-MDP with the AMF, we conducted coupled micromagnetic and mechanofluidic modeling. 
For this, two AF-coupled macroscopic magnetic moments $m_{1,2} = M_{\rm s}\cdot \pi r^2 t^{\rm m} = 1.675\cdot 10^{-15}$Am$^2$, with $M_{\rm s}^{\rm eff} = 1354\,$kA/m, $r=250\,$nm, and a ferromagnetic layer thickness $t^{\rm eff}_{\rm FL} = 6.3\,$nm were used. Further, a uniaxial anisotropy $K_{\rm u} = 20\,$kJ/m$^3$, an AF-coupling constant $J_{\rm RKKY} = t_{\rm s}^{\rm eff}\mu_0 M_{\rm s}^{\rm eff}H_{\rm ex}  = -0.17\,$mJ/m$^2$, with $\mu_0H_{\rm ex} = -20\,$mT and the viscosity of water $\eta = 0.89\cdot 10^{-3}\,$ Pa$\cdot$s at a temperature $T=25^{\circ}$C were employed. The obtained easy axis $M(H)$ loop (red curve in Fig.\,\ref{Fig:microfluidics_ac_align}{\bf a}) exhibits an abrupt switching between the AF and F states at fields comparable to those of the more gradual switching process observed in our experiment (see red $M(H)$-loop in Fig.\,\ref{Fig:results}{\bf a}). 

Our modeling then focused on a series of initial alignment angles between the easy axes of the SAF-MDP and the axis of the AMF with an amplitude of 40\,mT and a frequency of 305\,kHz, consistent with the parameters used in our experiments. We found that SAF-MDP with an initial easy-axis to AMF angle of 59$^{\circ}$, align their easy axis with the AMF within a few oscillation cycles (Fig.\,\ref{Fig:microfluidics_ac_align}{\bf b}) leading to an increasing hysteretic loss finally approaching that of the easy axis process (see $M(H)$-curves one to five in  Fig.\,\ref{Fig:microfluidics_ac_align}{\bf c}). For angles larger or equal to 60$^{\circ}$, the SAF-MDP align their hard axis with the AMF (Fig.\,\ref{Fig:microfluidics_ac_align}{\bf d}) and, for the case of our model, the hysteretic losses vanish (see non-hysteretic $M(H)$-curves one to five in  Fig.\,\ref{Fig:microfluidics_ac_align}{\bf e}). Our magneto-mechanofluid modeling thus reveals a critical angle of about 60$^{\circ}$, and that about 66\,\% of the SAF-MDP align their easy axis with the AMF and thus generate a hysteretic loss. This agrees reasonably well with our observation based on the experimental data which revealed that about 41.8\,\% of SAF-MDPs aligned their easy axes with the AMF. 

\section*{Conclusions and Perspectives}
Our research investigated the potential of designer nanoparticles, optimized through micromagnetic modeling, to achieve maximum specific loss power (SLP) based on fundamental physical principles.
 This evaluation was conducted under the constraint of maintaining patient comfort, defined by the biological discomfort level established by Hergt and Dutz, which sets the permissible maximum product of frequency and field amplitude.\,\cite{Hergt2007}. 
The holistic design approach presented here enables micromagnetic optimization of the particle design, subsequent particle fabrication using a scalable microfabrication approach, followed by in-depth magnetic characterization using in-vacuum, high-sensitivity MFM gaining insights into the magnetic switching behaviour at a single particle scale. These in-depth characterization in tandem with micromagnetic and mechanofluidic modeling enable unprecedented insights into the particle behavior and reveals that the presence of a a not yet perfect switching behavior between AF and F states as well as a yet partial alignment of the particle's easy axes with the AMF. 

Our findings further suggest that for the applied AMF with an amplitude $\mu_0H = \pm 40\,$mT, between 40-60\,\% of the particles align their hard axis to the AMF and consequently do not contribute to the hysteretic loss. Further simulations however suggested that a full alignment of the particle's easy axis with the AMF is possible either by the application of dc-field alignment pulses or an initially increased ac-field amplitude.

This work provides comprehensive mechanistic insights into the design and behavior of magnetic particles, significantly advancing our understanding in this field. By elucidating the fundamental principles that govern particle performance, this work opens new avenues for the creation of particles whose SLP is restricted only by the inherent laws of physics. This in-depth understanding effectively removes the traditional constraints imposed by sub-optimal particle properties due to material selection and fabrication limitations, thereby unleashing the potential for more efficient magnetic particle designs.

\bibliography{SAF-MDP}

\begin{thebibliography}{54}%
\makeatletter
\providecommand \@ifxundefined [1]{%
 \@ifx{#1\undefined}
}%
\providecommand \@ifnum [1]{%
 \ifnum #1\expandafter \@firstoftwo
 \else \expandafter \@secondoftwo
 \fi
}%
\providecommand \@ifx [1]{%
 \ifx #1\expandafter \@firstoftwo
 \else \expandafter \@secondoftwo
 \fi
}%
\providecommand \natexlab [1]{#1}%
\providecommand \enquote  [1]{``#1''}%
\providecommand \bibnamefont  [1]{#1}%
\providecommand \bibfnamefont [1]{#1}%
\providecommand \citenamefont [1]{#1}%
\providecommand \href@noop [0]{\@secondoftwo}%
\providecommand \href [0]{\begingroup \@sanitize@url \@href}%
\providecommand \@href[1]{\@@startlink{#1}\@@href}%
\providecommand \@@href[1]{\endgroup#1\@@endlink}%
\providecommand \@sanitize@url [0]{\catcode `\\12\catcode `\$12\catcode
  `\&12\catcode `\#12\catcode `\^12\catcode `\_12\catcode `\%12\relax}%
\providecommand \@@startlink[1]{}%
\providecommand \@@endlink[0]{}%
\providecommand \url  [0]{\begingroup\@sanitize@url \@url }%
\providecommand \@url [1]{\endgroup\@href {#1}{\urlprefix }}%
\providecommand \urlprefix  [0]{URL }%
\providecommand \Eprint [0]{\href }%
\providecommand \doibase [0]{https://doi.org/}%
\providecommand \selectlanguage [0]{\@gobble}%
\providecommand \bibinfo  [0]{\@secondoftwo}%
\providecommand \bibfield  [0]{\@secondoftwo}%
\providecommand \translation [1]{[#1]}%
\providecommand \BibitemOpen [0]{}%
\providecommand \bibitemStop [0]{}%
\providecommand \bibitemNoStop [0]{.\EOS\space}%
\providecommand \EOS [0]{\spacefactor3000\relax}%
\providecommand \BibitemShut  [1]{\csname bibitem#1\endcsname}%
\let\auto@bib@innerbib\@empty
\bibitem [{\citenamefont {Gavilán}\ \emph {et~al.}(2021)\citenamefont
  {Gavilán}, \citenamefont {Avugadda}, \citenamefont {Fernández-Cabada},
  \citenamefont {Soni}, \citenamefont {Cassani}, \citenamefont {Mai},
  \citenamefont {Chantrell},\ and\ \citenamefont {Pellegrino}}]{Gavilan2021}%
  \BibitemOpen
  \bibfield  {author} {\bibinfo {author} {\bibfnamefont {H.}~\bibnamefont
  {Gavilán}}, \bibinfo {author} {\bibfnamefont {S.~K.}\ \bibnamefont
  {Avugadda}}, \bibinfo {author} {\bibfnamefont {T.}~\bibnamefont
  {Fernández-Cabada}}, \bibinfo {author} {\bibfnamefont {N.}~\bibnamefont
  {Soni}}, \bibinfo {author} {\bibfnamefont {M.}~\bibnamefont {Cassani}},
  \bibinfo {author} {\bibfnamefont {B.~T.}\ \bibnamefont {Mai}}, \bibinfo
  {author} {\bibfnamefont {R.}~\bibnamefont {Chantrell}},\ and\ \bibinfo
  {author} {\bibfnamefont {T.}~\bibnamefont {Pellegrino}},\ }\bibfield  {title}
  {\bibinfo {title} {{Magnetic nanoparticles and clusters for magnetic
  hyperthermia: optimizing their heat performance and developing combinatorial
  therapies to tackle cancer}},\ }\href {https://doi.org/10.1039/d1cs00427a}
  {\bibfield  {journal} {\bibinfo  {journal} {Chemical Society Reviews}\
  }\textbf {\bibinfo {volume} {50}},\ \bibinfo {pages} {11614} (\bibinfo {year}
  {2021})}\BibitemShut {NoStop}%
\bibitem [{\citenamefont {Sharrock}(1994)}]{Sharrock1994}%
  \BibitemOpen
  \bibfield  {author} {\bibinfo {author} {\bibfnamefont {M.~P.}\ \bibnamefont
  {Sharrock}},\ }\bibfield  {title} {\bibinfo {title} {{Time dependence of
  switching fields in magnetic recording media (invited)}},\ }\href
  {https://doi.org/10.1063/1.358282} {\bibfield  {journal} {\bibinfo  {journal}
  {Journal of Applied Physics}\ }\textbf {\bibinfo {volume} {76}},\ \bibinfo
  {pages} {6413} (\bibinfo {year} {1994})}\BibitemShut {NoStop}%
\bibitem [{\citenamefont {Hergt}\ and\ \citenamefont {Dutz}(2007)}]{Hergt2007}%
  \BibitemOpen
  \bibfield  {author} {\bibinfo {author} {\bibfnamefont {R.}~\bibnamefont
  {Hergt}}\ and\ \bibinfo {author} {\bibfnamefont {S.}~\bibnamefont {Dutz}},\
  }\bibfield  {title} {\bibinfo {title} {{Magnetic particle
  hyperthermia—biophysical limitations of a visionary tumour therapy}},\
  }\href {https://doi.org/10.1016/j.jmmm.2006.10.1156} {\bibfield  {journal}
  {\bibinfo  {journal} {Journal Of Magnetism And Magnetic Materials}\ }\textbf
  {\bibinfo {volume} {311}},\ \bibinfo {pages} {187 } (\bibinfo {year}
  {2007})}\BibitemShut {NoStop}%
\bibitem [{\citenamefont {Helbig}\ \emph {et~al.}(2023)\citenamefont {Helbig},
  \citenamefont {Abert}, \citenamefont {Sánchez}, \citenamefont
  {Kantorovich},\ and\ \citenamefont {Suess}}]{Helbig2023}%
  \BibitemOpen
  \bibfield  {author} {\bibinfo {author} {\bibfnamefont {S.}~\bibnamefont
  {Helbig}}, \bibinfo {author} {\bibfnamefont {C.}~\bibnamefont {Abert}},
  \bibinfo {author} {\bibfnamefont {P.~A.}\ \bibnamefont {Sánchez}}, \bibinfo
  {author} {\bibfnamefont {S.~S.}\ \bibnamefont {Kantorovich}},\ and\ \bibinfo
  {author} {\bibfnamefont {D.}~\bibnamefont {Suess}},\ }\bibfield  {title}
  {\bibinfo {title} {{Self-consistent solution of magnetic and friction energy
  losses of a magnetic nanoparticle}},\ }\href
  {https://doi.org/10.1103/physrevb.107.054416} {\bibfield  {journal} {\bibinfo
   {journal} {Physical Review B}\ }\textbf {\bibinfo {volume} {107}},\ \bibinfo
  {pages} {054416} (\bibinfo {year} {2023})},\ \Eprint
  {https://arxiv.org/abs/2204.14106} {2204.14106} \BibitemShut {NoStop}%
\bibitem [{\citenamefont {Ruta}\ \emph {et~al.}(2015)\citenamefont {Ruta},
  \citenamefont {Chantrell},\ and\ \citenamefont {Hovorka}}]{Ruta2015}%
  \BibitemOpen
  \bibfield  {author} {\bibinfo {author} {\bibfnamefont {S.}~\bibnamefont
  {Ruta}}, \bibinfo {author} {\bibfnamefont {R.}~\bibnamefont {Chantrell}},\
  and\ \bibinfo {author} {\bibfnamefont {O.}~\bibnamefont {Hovorka}},\
  }\bibfield  {title} {\bibinfo {title} {{Unified model of hyperthermia via
  hysteresis heating in systems of interacting magnetic nanoparticles.}},\
  }\href {https://doi.org/10.1038/srep09090} {\bibfield  {journal} {\bibinfo
  {journal} {Scientific Reports}\ }\textbf {\bibinfo {volume} {5}},\ \bibinfo
  {pages} {9090} (\bibinfo {year} {2015})}\BibitemShut {NoStop}%
\bibitem [{\citenamefont {Thiesen}\ and\ \citenamefont
  {Jordan}(2008)}]{Thiesen2008}%
  \BibitemOpen
  \bibfield  {author} {\bibinfo {author} {\bibfnamefont {B.}~\bibnamefont
  {Thiesen}}\ and\ \bibinfo {author} {\bibfnamefont {A.}~\bibnamefont
  {Jordan}},\ }\bibfield  {title} {\bibinfo {title} {{Clinical applications of
  magnetic nanoparticles for hyperthermia}},\ }\href
  {https://doi.org/10.1080/02656730802104757} {\bibfield  {journal} {\bibinfo
  {journal} {International Journal of Hyperthermia}\ }\textbf {\bibinfo
  {volume} {24}},\ \bibinfo {pages} {467} (\bibinfo {year} {2008})}\BibitemShut
  {NoStop}%
\bibitem [{\citenamefont {Mamiya}(2013)}]{Mamiya2013}%
  \BibitemOpen
  \bibfield  {author} {\bibinfo {author} {\bibfnamefont {H.}~\bibnamefont
  {Mamiya}},\ }\bibfield  {title} {\bibinfo {title} {{Recent Advances in
  Understanding Magnetic Nanoparticles in AC Magnetic Fields and Optimal Design
  for Targeted Hyperthermia}},\ }\href {https://doi.org/10.1155/2013/752973}
  {\bibfield  {journal} {\bibinfo  {journal} {Journal of Nanomaterials}\
  }\textbf {\bibinfo {volume} {2013}},\ \bibinfo {pages} {1} (\bibinfo {year}
  {2013})}\BibitemShut {NoStop}%
\bibitem [{\citenamefont {Hu}\ \emph {et~al.}(2008)\citenamefont {Hu},
  \citenamefont {Wilson}, \citenamefont {Koh}, \citenamefont {Fu},
  \citenamefont {Faranesh}, \citenamefont {Earhart}, \citenamefont {Osterfeld},
  \citenamefont {Han}, \citenamefont {Xu}, \citenamefont {Guccione},
  \citenamefont {Sinclair},\ and\ \citenamefont {Wang}}]{Hu2008}%
  \BibitemOpen
  \bibfield  {author} {\bibinfo {author} {\bibfnamefont {W.}~\bibnamefont
  {Hu}}, \bibinfo {author} {\bibfnamefont {R.~J.}\ \bibnamefont {Wilson}},
  \bibinfo {author} {\bibfnamefont {A.}~\bibnamefont {Koh}}, \bibinfo {author}
  {\bibfnamefont {A.}~\bibnamefont {Fu}}, \bibinfo {author} {\bibfnamefont
  {A.~Z.}\ \bibnamefont {Faranesh}}, \bibinfo {author} {\bibfnamefont {C.~M.}\
  \bibnamefont {Earhart}}, \bibinfo {author} {\bibfnamefont {S.~J.}\
  \bibnamefont {Osterfeld}}, \bibinfo {author} {\bibfnamefont {S.-J.}\
  \bibnamefont {Han}}, \bibinfo {author} {\bibfnamefont {L.}~\bibnamefont
  {Xu}}, \bibinfo {author} {\bibfnamefont {S.}~\bibnamefont {Guccione}},
  \bibinfo {author} {\bibfnamefont {R.}~\bibnamefont {Sinclair}},\ and\
  \bibinfo {author} {\bibfnamefont {S.~X.}\ \bibnamefont {Wang}},\ }\bibfield
  {title} {\bibinfo {title} {{High-Moment Antiferromagnetic Nanoparticles with
  Tunable Magnetic Properties}},\ }\href
  {https://doi.org/10.1002/adma.200703077} {\bibfield  {journal} {\bibinfo
  {journal} {Advanced Materials}\ }\textbf {\bibinfo {volume} {20}},\ \bibinfo
  {pages} {1479 } (\bibinfo {year} {2008})}\BibitemShut {NoStop}%
\bibitem [{\citenamefont {Ruderman}\ and\ \citenamefont
  {Kittel}(1954)}]{Ruderman1954}%
  \BibitemOpen
  \bibfield  {author} {\bibinfo {author} {\bibfnamefont {M.~A.}\ \bibnamefont
  {Ruderman}}\ and\ \bibinfo {author} {\bibfnamefont {C.}~\bibnamefont
  {Kittel}},\ }\bibfield  {title} {\bibinfo {title} {{Indirect Exchange
  Coupling of Nuclear Magnetic Moments by Conduction Electrons}},\ }\href
  {https://doi.org/10.1103/physrev.96.99} {\bibfield  {journal} {\bibinfo
  {journal} {Physical Review}\ }\textbf {\bibinfo {volume} {96}},\ \bibinfo
  {pages} {99} (\bibinfo {year} {1954})}\BibitemShut {NoStop}%
\bibitem [{\citenamefont {Kasuya}(1956)}]{Kasuya1956}%
  \BibitemOpen
  \bibfield  {author} {\bibinfo {author} {\bibfnamefont {T.}~\bibnamefont
  {Kasuya}},\ }\bibfield  {title} {\bibinfo {title} {{A Theory of Metallic
  Ferro- and Antiferromagnetism on Zener's Model}},\ }\href
  {https://doi.org/10.1143/ptp.16.45} {\bibfield  {journal} {\bibinfo
  {journal} {Progress of Theoretical Physics}\ }\textbf {\bibinfo {volume}
  {16}},\ \bibinfo {pages} {45} (\bibinfo {year} {1956})}\BibitemShut {NoStop}%
\bibitem [{\citenamefont {Yosida}(1957)}]{Yosida1957}%
  \BibitemOpen
  \bibfield  {author} {\bibinfo {author} {\bibfnamefont {K.}~\bibnamefont
  {Yosida}},\ }\bibfield  {title} {\bibinfo {title} {{Magnetic Properties of
  Cu-Mn Alloys}},\ }\href {https://doi.org/10.1103/physrev.106.893} {\bibfield
  {journal} {\bibinfo  {journal} {Physical Review}\ }\textbf {\bibinfo {volume}
  {106}},\ \bibinfo {pages} {893} (\bibinfo {year} {1957})}\BibitemShut
  {NoStop}%
\bibitem [{\citenamefont {Stoner}\ and\ \citenamefont
  {Wohlfarth}(1948)}]{Stoner1948}%
  \BibitemOpen
  \bibfield  {author} {\bibinfo {author} {\bibfnamefont {E.~C.}\ \bibnamefont
  {Stoner}}\ and\ \bibinfo {author} {\bibfnamefont {E.~P.}\ \bibnamefont
  {Wohlfarth}},\ }\bibfield  {title} {\bibinfo {title} {{A mechanism of
  magnetic hysteresis in heterogeneous alloys}},\ }\href
  {https://doi.org/10.1098/rsta.1948.0007} {\bibfield  {journal} {\bibinfo
  {journal} {Philosophical Transactions of the Royal Society of London. Series
  A, Mathematical and Physical Sciences}\ }\textbf {\bibinfo {volume} {240}},\
  \bibinfo {pages} {599} (\bibinfo {year} {1948})}\BibitemShut {NoStop}%
\bibitem [{\citenamefont {Hu}\ \emph {et~al.}(2009)\citenamefont {Hu},
  \citenamefont {Wilson}, \citenamefont {Earhart}, \citenamefont {Koh},
  \citenamefont {Sinclair},\ and\ \citenamefont {Wang}}]{Hu2009}%
  \BibitemOpen
  \bibfield  {author} {\bibinfo {author} {\bibfnamefont {W.}~\bibnamefont
  {Hu}}, \bibinfo {author} {\bibfnamefont {R.~J.}\ \bibnamefont {Wilson}},
  \bibinfo {author} {\bibfnamefont {C.~M.}\ \bibnamefont {Earhart}}, \bibinfo
  {author} {\bibfnamefont {A.~L.}\ \bibnamefont {Koh}}, \bibinfo {author}
  {\bibfnamefont {R.}~\bibnamefont {Sinclair}},\ and\ \bibinfo {author}
  {\bibfnamefont {S.~X.}\ \bibnamefont {Wang}},\ }\bibfield  {title} {\bibinfo
  {title} {{Synthetic antiferromagnetic nanoparticles with tunable
  susceptibilities}},\ }\href {https://doi.org/10.1063/1.3072028} {\bibfield
  {journal} {\bibinfo  {journal} {Journal Of Applied Physics}\ }\textbf
  {\bibinfo {volume} {105}},\ \bibinfo {pages} {07B508 } (\bibinfo {year}
  {2009})}\BibitemShut {NoStop}%
\bibitem [{\citenamefont {Zhang}\ \emph {et~al.}(2013)\citenamefont {Zhang},
  \citenamefont {Earhart}, \citenamefont {Ooi}, \citenamefont {Wilson},
  \citenamefont {Tang},\ and\ \citenamefont {Wang}}]{Zhang2013}%
  \BibitemOpen
  \bibfield  {author} {\bibinfo {author} {\bibfnamefont {M.}~\bibnamefont
  {Zhang}}, \bibinfo {author} {\bibfnamefont {C.~M.}\ \bibnamefont {Earhart}},
  \bibinfo {author} {\bibfnamefont {C.}~\bibnamefont {Ooi}}, \bibinfo {author}
  {\bibfnamefont {R.~J.}\ \bibnamefont {Wilson}}, \bibinfo {author}
  {\bibfnamefont {M.}~\bibnamefont {Tang}},\ and\ \bibinfo {author}
  {\bibfnamefont {S.~X.}\ \bibnamefont {Wang}},\ }\bibfield  {title} {\bibinfo
  {title} {{Functionalization of high-moment magnetic nanodisks for cell
  manipulation and separation}},\ }\href
  {https://doi.org/10.1007/s12274-013-0352-4} {\bibfield  {journal} {\bibinfo
  {journal} {Nano Research}\ }\textbf {\bibinfo {volume} {6}},\ \bibinfo
  {pages} {745 } (\bibinfo {year} {2013})}\BibitemShut {NoStop}%
\bibitem [{\citenamefont {Vemulkar}\ \emph {et~al.}(2015)\citenamefont
  {Vemulkar}, \citenamefont {Mansell}, \citenamefont {Petit}, \citenamefont
  {Cowburn},\ and\ \citenamefont {Lesniak}}]{Vemulkar2015}%
  \BibitemOpen
  \bibfield  {author} {\bibinfo {author} {\bibfnamefont {T.}~\bibnamefont
  {Vemulkar}}, \bibinfo {author} {\bibfnamefont {R.}~\bibnamefont {Mansell}},
  \bibinfo {author} {\bibfnamefont {D.~C. M.~C.}\ \bibnamefont {Petit}},
  \bibinfo {author} {\bibfnamefont {R.~P.}\ \bibnamefont {Cowburn}},\ and\
  \bibinfo {author} {\bibfnamefont {M.~S.}\ \bibnamefont {Lesniak}},\
  }\bibfield  {title} {\bibinfo {title} {{Highly tunable perpendicularly
  magnetized synthetic antiferromagnets for biotechnology applications}},\
  }\href {https://doi.org/10.1063/1.4926336} {\bibfield  {journal} {\bibinfo
  {journal} {Applied Physics Letters}\ }\textbf {\bibinfo {volume} {107}},\
  \bibinfo {pages} {012403 } (\bibinfo {year} {2015})}\BibitemShut {NoStop}%
\bibitem [{\citenamefont {Vemulkar}\ \emph {et~al.}(2017)\citenamefont
  {Vemulkar}, \citenamefont {Welbourne}, \citenamefont {Mansell}, \citenamefont
  {Petit},\ and\ \citenamefont {Cowburn}}]{Vemulkar2017}%
  \BibitemOpen
  \bibfield  {author} {\bibinfo {author} {\bibfnamefont {T.}~\bibnamefont
  {Vemulkar}}, \bibinfo {author} {\bibfnamefont {E.~N.}\ \bibnamefont
  {Welbourne}}, \bibinfo {author} {\bibfnamefont {R.}~\bibnamefont {Mansell}},
  \bibinfo {author} {\bibfnamefont {D.~C. M.~C.}\ \bibnamefont {Petit}},\ and\
  \bibinfo {author} {\bibfnamefont {R.~P.}\ \bibnamefont {Cowburn}},\
  }\bibfield  {title} {\bibinfo {title} {{The mechanical response in a fluid of
  synthetic antiferromagnetic and ferrimagnetic microdiscs with perpendicular
  magnetic anisotropy}},\ }\href {https://doi.org/10.1063/1.4974211} {\bibfield
   {journal} {\bibinfo  {journal} {Applied Physics Letters}\ }\textbf {\bibinfo
  {volume} {110}},\ \bibinfo {pages} {042402} (\bibinfo {year}
  {2017})}\BibitemShut {NoStop}%
\bibitem [{\citenamefont {Cheng}\ \emph {et~al.}(2016)\citenamefont {Cheng},
  \citenamefont {Muroski}, \citenamefont {Petit}, \citenamefont {Mansell},
  \citenamefont {Vemulkar}, \citenamefont {Morshed}, \citenamefont {Han},
  \citenamefont {Balyasnikova}, \citenamefont {Horbinski}, \citenamefont
  {Huang}, \citenamefont {Zhang}, \citenamefont {Cowburn},\ and\ \citenamefont
  {Lesniak}}]{Cheng2016}%
  \BibitemOpen
  \bibfield  {author} {\bibinfo {author} {\bibfnamefont {Y.}~\bibnamefont
  {Cheng}}, \bibinfo {author} {\bibfnamefont {M.~E.}\ \bibnamefont {Muroski}},
  \bibinfo {author} {\bibfnamefont {D.~C.}\ \bibnamefont {Petit}}, \bibinfo
  {author} {\bibfnamefont {R.}~\bibnamefont {Mansell}}, \bibinfo {author}
  {\bibfnamefont {T.}~\bibnamefont {Vemulkar}}, \bibinfo {author}
  {\bibfnamefont {R.~A.}\ \bibnamefont {Morshed}}, \bibinfo {author}
  {\bibfnamefont {Y.}~\bibnamefont {Han}}, \bibinfo {author} {\bibfnamefont
  {I.~V.}\ \bibnamefont {Balyasnikova}}, \bibinfo {author} {\bibfnamefont
  {C.~M.}\ \bibnamefont {Horbinski}}, \bibinfo {author} {\bibfnamefont
  {X.}~\bibnamefont {Huang}}, \bibinfo {author} {\bibfnamefont
  {L.}~\bibnamefont {Zhang}}, \bibinfo {author} {\bibfnamefont {R.~P.}\
  \bibnamefont {Cowburn}},\ and\ \bibinfo {author} {\bibfnamefont {M.~S.}\
  \bibnamefont {Lesniak}},\ }\bibfield  {title} {\bibinfo {title} {{Rotating
  magnetic field induced oscillation of magnetic particles for in vivo
  mechanical destruction of malignant glioma}},\ }\href
  {https://doi.org/10.1016/j.jconrel.2015.12.028} {\bibfield  {journal}
  {\bibinfo  {journal} {Journal of Controlled Release}\ }\textbf {\bibinfo
  {volume} {223}},\ \bibinfo {pages} {75} (\bibinfo {year} {2016})}\BibitemShut
  {NoStop}%
\bibitem [{\citenamefont {Mansell}\ \emph {et~al.}(2017)\citenamefont
  {Mansell}, \citenamefont {Vemulkar}, \citenamefont {Petit}, \citenamefont
  {Cheng}, \citenamefont {Murphy}, \citenamefont {Lesniak},\ and\ \citenamefont
  {Cowburn}}]{Mansell2017}%
  \BibitemOpen
  \bibfield  {author} {\bibinfo {author} {\bibfnamefont {R.}~\bibnamefont
  {Mansell}}, \bibinfo {author} {\bibfnamefont {T.}~\bibnamefont {Vemulkar}},
  \bibinfo {author} {\bibfnamefont {D.~C. M.~C.}\ \bibnamefont {Petit}},
  \bibinfo {author} {\bibfnamefont {Y.}~\bibnamefont {Cheng}}, \bibinfo
  {author} {\bibfnamefont {J.}~\bibnamefont {Murphy}}, \bibinfo {author}
  {\bibfnamefont {M.~S.}\ \bibnamefont {Lesniak}},\ and\ \bibinfo {author}
  {\bibfnamefont {R.~P.}\ \bibnamefont {Cowburn}},\ }\bibfield  {title}
  {\bibinfo {title} {{Magnetic particles with perpendicular anisotropy for
  mechanical cancer cell destruction}},\ }\href
  {https://doi.org/10.1038/s41598-017-04154-1} {\bibfield  {journal} {\bibinfo
  {journal} {Scientific Reports}\ }\textbf {\bibinfo {volume} {7}},\ \bibinfo
  {pages} {4257} (\bibinfo {year} {2017})}\BibitemShut {NoStop}%
\bibitem [{\citenamefont {Kim}\ \emph {et~al.}(2010)\citenamefont {Kim},
  \citenamefont {Rozhkova}, \citenamefont {Ulasov}, \citenamefont {Bader},
  \citenamefont {Rajh}, \citenamefont {Lesniak},\ and\ \citenamefont
  {Novosad}}]{Kim2010}%
  \BibitemOpen
  \bibfield  {author} {\bibinfo {author} {\bibfnamefont {D.-H.}\ \bibnamefont
  {Kim}}, \bibinfo {author} {\bibfnamefont {E.~A.}\ \bibnamefont {Rozhkova}},
  \bibinfo {author} {\bibfnamefont {I.~V.}\ \bibnamefont {Ulasov}}, \bibinfo
  {author} {\bibfnamefont {S.~D.}\ \bibnamefont {Bader}}, \bibinfo {author}
  {\bibfnamefont {T.}~\bibnamefont {Rajh}}, \bibinfo {author} {\bibfnamefont
  {M.~S.}\ \bibnamefont {Lesniak}},\ and\ \bibinfo {author} {\bibfnamefont
  {V.}~\bibnamefont {Novosad}},\ }\bibfield  {title} {\bibinfo {title}
  {{Biofunctionalized magnetic-vortex microdiscs for targeted cancer-cell
  destruction}},\ }\href {https://doi.org/10.1038/nmat2591} {\bibfield
  {journal} {\bibinfo  {journal} {Nature Materials}\ }\textbf {\bibinfo
  {volume} {9}},\ \bibinfo {pages} {165} (\bibinfo {year} {2010})}\BibitemShut
  {NoStop}%
\bibitem [{\citenamefont {Leulmi}\ \emph {et~al.}(2015)\citenamefont {Leulmi},
  \citenamefont {Chauchet}, \citenamefont {Morcrette}, \citenamefont {Ortiz},
  \citenamefont {Joisten}, \citenamefont {Sabon}, \citenamefont {Livache},
  \citenamefont {Hou}, \citenamefont {Carrière}, \citenamefont {Lequien},\
  and\ \citenamefont {Dieny}}]{Leulmi2015}%
  \BibitemOpen
  \bibfield  {author} {\bibinfo {author} {\bibfnamefont {S.}~\bibnamefont
  {Leulmi}}, \bibinfo {author} {\bibfnamefont {X.}~\bibnamefont {Chauchet}},
  \bibinfo {author} {\bibfnamefont {M.}~\bibnamefont {Morcrette}}, \bibinfo
  {author} {\bibfnamefont {G.}~\bibnamefont {Ortiz}}, \bibinfo {author}
  {\bibfnamefont {H.}~\bibnamefont {Joisten}}, \bibinfo {author} {\bibfnamefont
  {P.}~\bibnamefont {Sabon}}, \bibinfo {author} {\bibfnamefont
  {T.}~\bibnamefont {Livache}}, \bibinfo {author} {\bibfnamefont
  {Y.}~\bibnamefont {Hou}}, \bibinfo {author} {\bibfnamefont {M.}~\bibnamefont
  {Carrière}}, \bibinfo {author} {\bibfnamefont {S.}~\bibnamefont {Lequien}},\
  and\ \bibinfo {author} {\bibfnamefont {B.}~\bibnamefont {Dieny}},\ }\bibfield
   {title} {\bibinfo {title} {{Triggering the apoptosis of targeted human renal
  cancer cells by the vibration of anisotropic magnetic particles attached to
  the cell membrane}},\ }\href {https://doi.org/10.1039/c5nr03518j} {\bibfield
  {journal} {\bibinfo  {journal} {Nanoscale}\ }\textbf {\bibinfo {volume}
  {7}},\ \bibinfo {pages} {15904 } (\bibinfo {year} {2015})}\BibitemShut
  {NoStop}%
\bibitem [{\citenamefont {Goiriena-Goikoetxea}\ \emph
  {et~al.}(2020)\citenamefont {Goiriena-Goikoetxea}, \citenamefont {Muñoz},
  \citenamefont {Orue}, \citenamefont {Fernández-Gubieda}, \citenamefont
  {Bokor}, \citenamefont {Muela},\ and\ \citenamefont
  {García-Arribas}}]{GoirienaGoikoetxea2020}%
  \BibitemOpen
  \bibfield  {author} {\bibinfo {author} {\bibfnamefont {M.}~\bibnamefont
  {Goiriena-Goikoetxea}}, \bibinfo {author} {\bibfnamefont {D.}~\bibnamefont
  {Muñoz}}, \bibinfo {author} {\bibfnamefont {I.}~\bibnamefont {Orue}},
  \bibinfo {author} {\bibfnamefont {M.~L.}\ \bibnamefont {Fernández-Gubieda}},
  \bibinfo {author} {\bibfnamefont {J.}~\bibnamefont {Bokor}}, \bibinfo
  {author} {\bibfnamefont {A.}~\bibnamefont {Muela}},\ and\ \bibinfo {author}
  {\bibfnamefont {A.}~\bibnamefont {García-Arribas}},\ }\bibfield  {title}
  {\bibinfo {title} {{Disk-shaped magnetic particles for cancer therapy}},\
  }\href {https://doi.org/10.1063/1.5123716} {\bibfield  {journal} {\bibinfo
  {journal} {Applied Physics Reviews}\ }\textbf {\bibinfo {volume} {7}},\
  \bibinfo {pages} {011306} (\bibinfo {year} {2020})}\BibitemShut {NoStop}%
\bibitem [{\citenamefont {Naud}\ \emph {et~al.}(2020)\citenamefont {Naud},
  \citenamefont {Thébault}, \citenamefont {Carrière}, \citenamefont {Hou},
  \citenamefont {Morel}, \citenamefont {Berger}, \citenamefont {Diény},\ and\
  \citenamefont {Joisten}}]{Naud2020}%
  \BibitemOpen
  \bibfield  {author} {\bibinfo {author} {\bibfnamefont {C.}~\bibnamefont
  {Naud}}, \bibinfo {author} {\bibfnamefont {C.}~\bibnamefont {Thébault}},
  \bibinfo {author} {\bibfnamefont {M.}~\bibnamefont {Carrière}}, \bibinfo
  {author} {\bibfnamefont {Y.}~\bibnamefont {Hou}}, \bibinfo {author}
  {\bibfnamefont {R.}~\bibnamefont {Morel}}, \bibinfo {author} {\bibfnamefont
  {F.}~\bibnamefont {Berger}}, \bibinfo {author} {\bibfnamefont
  {B.}~\bibnamefont {Diény}},\ and\ \bibinfo {author} {\bibfnamefont
  {H.}~\bibnamefont {Joisten}},\ }\bibfield  {title} {\bibinfo {title} {{Cancer
  treatment by magneto-mechanical effect of particles, a review}},\ }\href
  {https://doi.org/10.1039/d0na00187b} {\bibfield  {journal} {\bibinfo
  {journal} {Nanoscale Advances}\ }\textbf {\bibinfo {volume} {2}},\ \bibinfo
  {pages} {3632} (\bibinfo {year} {2020})}\BibitemShut {NoStop}%
\bibitem [{\citenamefont {Li}\ \emph {et~al.}(2022)\citenamefont {Li},
  \citenamefont {Nieuwkerk}, \citenamefont {Verschuuren}, \citenamefont
  {Koopmans},\ and\ \citenamefont {Lavrijsen}}]{Li2022}%
  \BibitemOpen
  \bibfield  {author} {\bibinfo {author} {\bibfnamefont {J.}~\bibnamefont
  {Li}}, \bibinfo {author} {\bibfnamefont {P.~v.}\ \bibnamefont {Nieuwkerk}},
  \bibinfo {author} {\bibfnamefont {M.~A.}\ \bibnamefont {Verschuuren}},
  \bibinfo {author} {\bibfnamefont {B.}~\bibnamefont {Koopmans}},\ and\
  \bibinfo {author} {\bibfnamefont {R.}~\bibnamefont {Lavrijsen}},\ }\bibfield
  {title} {\bibinfo {title} {{Substrate conformal imprint fabrication process
  of synthetic antiferromagnetic nanoplatelets}},\ }\href
  {https://doi.org/10.1063/5.0100657} {\bibfield  {journal} {\bibinfo
  {journal} {Applied Physics Letters}\ }\textbf {\bibinfo {volume} {121}},\
  \bibinfo {pages} {182407} (\bibinfo {year} {2022})},\ \Eprint
  {https://arxiv.org/abs/2206.15320} {2206.15320} \BibitemShut {NoStop}%
\bibitem [{\citenamefont {Adhikari}\ \emph {et~al.}(2023)\citenamefont
  {Adhikari}, \citenamefont {Li}, \citenamefont {Wang}, \citenamefont {Ruijs},
  \citenamefont {Liu}, \citenamefont {Koopmans}, \citenamefont {Orrit},\ and\
  \citenamefont {Lavrijsen}}]{Adhikari2023}%
  \BibitemOpen
  \bibfield  {author} {\bibinfo {author} {\bibfnamefont {S.}~\bibnamefont
  {Adhikari}}, \bibinfo {author} {\bibfnamefont {J.}~\bibnamefont {Li}},
  \bibinfo {author} {\bibfnamefont {Y.}~\bibnamefont {Wang}}, \bibinfo {author}
  {\bibfnamefont {L.}~\bibnamefont {Ruijs}}, \bibinfo {author} {\bibfnamefont
  {J.}~\bibnamefont {Liu}}, \bibinfo {author} {\bibfnamefont {B.}~\bibnamefont
  {Koopmans}}, \bibinfo {author} {\bibfnamefont {M.}~\bibnamefont {Orrit}},\
  and\ \bibinfo {author} {\bibfnamefont {R.}~\bibnamefont {Lavrijsen}},\
  }\bibfield  {title} {\bibinfo {title} {{Optical Monitoring of the
  Magnetization Switching of Single Synthetic-Antiferromagnetic Nanoplatelets
  with Perpendicular Magnetic Anisotropy}},\ }\href
  {https://doi.org/10.1021/acsphotonics.3c00123} {\bibfield  {journal}
  {\bibinfo  {journal} {ACS Photonics}\ }\textbf {\bibinfo {volume} {10}},\
  \bibinfo {pages} {1512} (\bibinfo {year} {2023})}\BibitemShut {NoStop}%
\bibitem [{\citenamefont {Jamet}\ \emph {et~al.}(1998)\citenamefont {Jamet},
  \citenamefont {Lemerle}, \citenamefont {Meyer}, \citenamefont {Ferré},
  \citenamefont {Bartenlian}, \citenamefont {Bardou}, \citenamefont {Chappert},
  \citenamefont {Veillet}, \citenamefont {Rousseaux}, \citenamefont
  {Decanini},\ and\ \citenamefont {Launois}}]{Jamet1998}%
  \BibitemOpen
  \bibfield  {author} {\bibinfo {author} {\bibfnamefont {J.-P.}\ \bibnamefont
  {Jamet}}, \bibinfo {author} {\bibfnamefont {S.}~\bibnamefont {Lemerle}},
  \bibinfo {author} {\bibfnamefont {P.}~\bibnamefont {Meyer}}, \bibinfo
  {author} {\bibfnamefont {J.}~\bibnamefont {Ferré}}, \bibinfo {author}
  {\bibfnamefont {B.}~\bibnamefont {Bartenlian}}, \bibinfo {author}
  {\bibfnamefont {N.}~\bibnamefont {Bardou}}, \bibinfo {author} {\bibfnamefont
  {C.}~\bibnamefont {Chappert}}, \bibinfo {author} {\bibfnamefont
  {P.}~\bibnamefont {Veillet}}, \bibinfo {author} {\bibfnamefont
  {F.}~\bibnamefont {Rousseaux}}, \bibinfo {author} {\bibfnamefont
  {D.}~\bibnamefont {Decanini}},\ and\ \bibinfo {author} {\bibfnamefont
  {H.}~\bibnamefont {Launois}},\ }\bibfield  {title} {\bibinfo {title}
  {{Dynamics of the magnetization reversal in Au/Co/Au micrometer-size dot
  arrays}},\ }\href {https://doi.org/10.1103/physrevb.57.14320} {\bibfield
  {journal} {\bibinfo  {journal} {Physical Review B}\ }\textbf {\bibinfo
  {volume} {57}},\ \bibinfo {pages} {14320} (\bibinfo {year}
  {1998})}\BibitemShut {NoStop}%
\bibitem [{\citenamefont {Thomson}\ \emph {et~al.}(2006)\citenamefont
  {Thomson}, \citenamefont {Hu},\ and\ \citenamefont {Terris}}]{Thomson2006}%
  \BibitemOpen
  \bibfield  {author} {\bibinfo {author} {\bibfnamefont {T.}~\bibnamefont
  {Thomson}}, \bibinfo {author} {\bibfnamefont {G.}~\bibnamefont {Hu}},\ and\
  \bibinfo {author} {\bibfnamefont {B.}~\bibnamefont {Terris}},\ }\bibfield
  {title} {\bibinfo {title} {{Intrinsic Distribution of Magnetic Anisotropy in
  Thin Films Probed by Patterned Nanostructures}}\ }\textbf {\bibinfo {volume}
  {96}},\ \href {https://doi.org/10.1103/physrevlett.96.257204}
  {10.1103/physrevlett.96.257204} (\bibinfo {year} {2006})\BibitemShut
  {NoStop}%
\bibitem [{\citenamefont {Shaw}\ \emph {et~al.}(2007)\citenamefont {Shaw},
  \citenamefont {Rippard}, \citenamefont {Russek}, \citenamefont {Reith},\ and\
  \citenamefont {Falco}}]{Shaw2007}%
  \BibitemOpen
  \bibfield  {author} {\bibinfo {author} {\bibfnamefont {J.~M.}\ \bibnamefont
  {Shaw}}, \bibinfo {author} {\bibfnamefont {W.~H.}\ \bibnamefont {Rippard}},
  \bibinfo {author} {\bibfnamefont {S.~E.}\ \bibnamefont {Russek}}, \bibinfo
  {author} {\bibfnamefont {T.}~\bibnamefont {Reith}},\ and\ \bibinfo {author}
  {\bibfnamefont {C.~M.}\ \bibnamefont {Falco}},\ }\bibfield  {title} {\bibinfo
  {title} {{Origins of switching field distributions in perpendicular magnetic
  nanodot arrays}},\ }\href {https://doi.org/10.1063/1.2431399} {\bibfield
  {journal} {\bibinfo  {journal} {Journal of Applied Physics}\ }\textbf
  {\bibinfo {volume} {101}},\ \bibinfo {pages} {023909} (\bibinfo {year}
  {2007})}\BibitemShut {NoStop}%
\bibitem [{\citenamefont {Pfau}\ \emph {et~al.}(2011)\citenamefont {Pfau},
  \citenamefont {Günther}, \citenamefont {Guehrs}, \citenamefont {Hauet},
  \citenamefont {Yang}, \citenamefont {Vinh}, \citenamefont {Xu}, \citenamefont
  {Yaney}, \citenamefont {Rick}, \citenamefont {Eisebitt},\ and\ \citenamefont
  {Hellwig}}]{Pfau2011}%
  \BibitemOpen
  \bibfield  {author} {\bibinfo {author} {\bibfnamefont {B.}~\bibnamefont
  {Pfau}}, \bibinfo {author} {\bibfnamefont {C.~M.}\ \bibnamefont {Günther}},
  \bibinfo {author} {\bibfnamefont {E.}~\bibnamefont {Guehrs}}, \bibinfo
  {author} {\bibfnamefont {T.}~\bibnamefont {Hauet}}, \bibinfo {author}
  {\bibfnamefont {H.}~\bibnamefont {Yang}}, \bibinfo {author} {\bibfnamefont
  {L.}~\bibnamefont {Vinh}}, \bibinfo {author} {\bibfnamefont {X.}~\bibnamefont
  {Xu}}, \bibinfo {author} {\bibfnamefont {D.}~\bibnamefont {Yaney}}, \bibinfo
  {author} {\bibfnamefont {R.}~\bibnamefont {Rick}}, \bibinfo {author}
  {\bibfnamefont {S.}~\bibnamefont {Eisebitt}},\ and\ \bibinfo {author}
  {\bibfnamefont {O.}~\bibnamefont {Hellwig}},\ }\bibfield  {title} {\bibinfo
  {title} {{Origin of magnetic switching field distribution in bit patterned
  media based on pre-patterned substrates}},\ }\href
  {https://doi.org/10.1063/1.3623488} {\bibfield  {journal} {\bibinfo
  {journal} {Applied Physics Letters}\ }\textbf {\bibinfo {volume} {99}},\
  \bibinfo {pages} {062502} (\bibinfo {year} {2011})}\BibitemShut {NoStop}%
\bibitem [{\citenamefont {Liu}\ and\ \citenamefont {Morisako}(2008)}]{Liu2008}%
  \BibitemOpen
  \bibfield  {author} {\bibinfo {author} {\bibfnamefont {X.}~\bibnamefont
  {Liu}}\ and\ \bibinfo {author} {\bibfnamefont {A.}~\bibnamefont {Morisako}},\
  }\bibfield  {title} {\bibinfo {title} {{Magnetic Properties of FeCo Films
  Prepared by Co-Sputtering and Hydrogenous Gas Reactive Sputtering}},\ }\href
  {https://doi.org/10.1109/tmag.2008.2001349} {\bibfield  {journal} {\bibinfo
  {journal} {IEEE Transactions on Magnetics}\ }\textbf {\bibinfo {volume}
  {44}},\ \bibinfo {pages} {3910} (\bibinfo {year} {2008})}\BibitemShut
  {NoStop}%
\bibitem [{\citenamefont {Suess}\ \emph {et~al.}(2016)\citenamefont {Suess},
  \citenamefont {Fuger}, \citenamefont {Abert}, \citenamefont {Bruckner},\ and\
  \citenamefont {Vogler}}]{suess2016superior}%
  \BibitemOpen
  \bibfield  {author} {\bibinfo {author} {\bibfnamefont {D.}~\bibnamefont
  {Suess}}, \bibinfo {author} {\bibfnamefont {M.}~\bibnamefont {Fuger}},
  \bibinfo {author} {\bibfnamefont {C.}~\bibnamefont {Abert}}, \bibinfo
  {author} {\bibfnamefont {F.}~\bibnamefont {Bruckner}},\ and\ \bibinfo
  {author} {\bibfnamefont {C.}~\bibnamefont {Vogler}},\ }\bibfield  {title}
  {\bibinfo {title} {Superior bit error rate and jitter due to improved
  switching field distribution in exchange spring magnetic recording media},\
  }\href@noop {} {\bibfield  {journal} {\bibinfo  {journal} {Scientific
  Reports}\ }\textbf {\bibinfo {volume} {6}},\ \bibinfo {pages} {27048}
  (\bibinfo {year} {2016})}\BibitemShut {NoStop}%
\bibitem [{\citenamefont {Cowburn}\ \emph {et~al.}(1999)\citenamefont
  {Cowburn}, \citenamefont {Koltsov}, \citenamefont {Adeyeye}, \citenamefont
  {Welland},\ and\ \citenamefont {Tricker}}]{Cowburn1999}%
  \BibitemOpen
  \bibfield  {author} {\bibinfo {author} {\bibfnamefont {R.~P.}\ \bibnamefont
  {Cowburn}}, \bibinfo {author} {\bibfnamefont {D.~K.}\ \bibnamefont
  {Koltsov}}, \bibinfo {author} {\bibfnamefont {A.~O.}\ \bibnamefont
  {Adeyeye}}, \bibinfo {author} {\bibfnamefont {M.~E.}\ \bibnamefont
  {Welland}},\ and\ \bibinfo {author} {\bibfnamefont {D.~M.}\ \bibnamefont
  {Tricker}},\ }\bibfield  {title} {\bibinfo {title} {{Single-Domain Circular
  Nanomagnets}},\ }\href {https://doi.org/10.1103/physrevlett.83.1042}
  {\bibfield  {journal} {\bibinfo  {journal} {Physical Review Letters}\
  }\textbf {\bibinfo {volume} {83}},\ \bibinfo {pages} {1042} (\bibinfo {year}
  {1999})}\BibitemShut {NoStop}%
\bibitem [{\citenamefont {Usov}\ and\ \citenamefont
  {Peschany}(1993)}]{Usov1993}%
  \BibitemOpen
  \bibfield  {author} {\bibinfo {author} {\bibfnamefont {N.}~\bibnamefont
  {Usov}}\ and\ \bibinfo {author} {\bibfnamefont {S.}~\bibnamefont
  {Peschany}},\ }\bibfield  {title} {\bibinfo {title} {{Magnetization curling
  in a fine cylindrical particle}},\ }\href
  {https://doi.org/10.1016/0304-8853(93)90428-5} {\bibfield  {journal}
  {\bibinfo  {journal} {Journal of Magnetism and Magnetic Materials}\ }\textbf
  {\bibinfo {volume} {118}},\ \bibinfo {pages} {L290} (\bibinfo {year}
  {1993})}\BibitemShut {NoStop}%
\bibitem [{\citenamefont {Guslienko}\ and\ \citenamefont
  {Metlov}(2001)}]{Guslienko2001}%
  \BibitemOpen
  \bibfield  {author} {\bibinfo {author} {\bibfnamefont {K.~Y.}\ \bibnamefont
  {Guslienko}}\ and\ \bibinfo {author} {\bibfnamefont {K.~L.}\ \bibnamefont
  {Metlov}},\ }\bibfield  {title} {\bibinfo {title} {{Evolution and stability
  of a magnetic vortex in a small cylindrical ferromagnetic particle under
  applied field}},\ }\href {https://doi.org/10.1103/physrevb.63.100403}
  {\bibfield  {journal} {\bibinfo  {journal} {Physical Review B}\ }\textbf
  {\bibinfo {volume} {63}},\ \bibinfo {pages} {100403} (\bibinfo {year}
  {2001})},\ \Eprint {https://arxiv.org/abs/cond-mat/0012299}
  {cond-mat/0012299} \BibitemShut {NoStop}%
\bibitem [{\citenamefont {Scheibler}\ \emph {et~al.}(2023)\citenamefont
  {Scheibler}, \citenamefont {Yildirim}, \citenamefont {Herrmann},\ and\
  \citenamefont {Hug}}]{Scheibler2023}%
  \BibitemOpen
  \bibfield  {author} {\bibinfo {author} {\bibfnamefont {S.}~\bibnamefont
  {Scheibler}}, \bibinfo {author} {\bibfnamefont {O.}~\bibnamefont {Yildirim}},
  \bibinfo {author} {\bibfnamefont {I.}~\bibnamefont {Herrmann}},\ and\
  \bibinfo {author} {\bibfnamefont {H.}~\bibnamefont {Hug}},\ }\bibfield
  {title} {\bibinfo {title} {{Inducing in-plane uniaxial magnetic anisotropies
  in amorphous CoFeB thin films}},\ }\href
  {https://doi.org/10.1016/j.jmmm.2023.171015} {\bibfield  {journal} {\bibinfo
  {journal} {Journal of Magnetism and Magnetic Materials}\ }\textbf {\bibinfo
  {volume} {585}},\ \bibinfo {pages} {171015} (\bibinfo {year}
  {2023})}\BibitemShut {NoStop}%
\bibitem [{\citenamefont {Magnus}\ \emph {et~al.}(2013)\citenamefont {Magnus},
  \citenamefont {Moubah}, \citenamefont {Roos}, \citenamefont {Kruk},
  \citenamefont {Kapaklis}, \citenamefont {Hase}, \citenamefont
  {Hjörvarsson},\ and\ \citenamefont {Andersson}}]{Magnus2013}%
  \BibitemOpen
  \bibfield  {author} {\bibinfo {author} {\bibfnamefont {F.}~\bibnamefont
  {Magnus}}, \bibinfo {author} {\bibfnamefont {R.}~\bibnamefont {Moubah}},
  \bibinfo {author} {\bibfnamefont {A.~H.}\ \bibnamefont {Roos}}, \bibinfo
  {author} {\bibfnamefont {A.}~\bibnamefont {Kruk}}, \bibinfo {author}
  {\bibfnamefont {V.}~\bibnamefont {Kapaklis}}, \bibinfo {author}
  {\bibfnamefont {T.}~\bibnamefont {Hase}}, \bibinfo {author} {\bibfnamefont
  {B.}~\bibnamefont {Hjörvarsson}},\ and\ \bibinfo {author} {\bibfnamefont
  {G.}~\bibnamefont {Andersson}},\ }\bibfield  {title} {\bibinfo {title}
  {{Tunable giant magnetic anisotropy in amorphous SmCo thin films}},\
  }\bibfield  {journal} {\bibinfo  {journal} {Applied Physics Letters}\
  }\textbf {\bibinfo {volume} {102}},\ \href
  {https://doi.org/10.1063/1.4802908} {10.1063/1.4802908} (\bibinfo {year}
  {2013})\BibitemShut {NoStop}%
\bibitem [{\citenamefont {Rybczynski}\ \emph {et~al.}(2003)\citenamefont
  {Rybczynski}, \citenamefont {Ebels},\ and\ \citenamefont
  {Giersig}}]{Rybczynski2003}%
  \BibitemOpen
  \bibfield  {author} {\bibinfo {author} {\bibfnamefont {J.}~\bibnamefont
  {Rybczynski}}, \bibinfo {author} {\bibfnamefont {U.}~\bibnamefont {Ebels}},\
  and\ \bibinfo {author} {\bibfnamefont {M.}~\bibnamefont {Giersig}},\
  }\bibfield  {title} {\bibinfo {title} {{Large-scale, 2D arrays of magnetic
  nanoparticles}},\ }\href {https://doi.org/10.1016/S0927-7757(03)00011-6}
  {\bibfield  {journal} {\bibinfo  {journal} {Colloids and Surfaces A:
  Physicochem. Eng. Aspects}\ }\textbf {\bibinfo {volume} {219}},\ \bibinfo
  {pages} {1} (\bibinfo {year} {2003})}\BibitemShut {NoStop}%
\bibitem [{\citenamefont {Armstrong}\ and\ \citenamefont
  {O'Dwyer}(2015)}]{ArmstrongEileen2015}%
  \BibitemOpen
  \bibfield  {author} {\bibinfo {author} {\bibfnamefont {E.}~\bibnamefont
  {Armstrong}}\ and\ \bibinfo {author} {\bibfnamefont {C.}~\bibnamefont
  {O'Dwyer}},\ }\bibfield  {title} {\bibinfo {title} {{Artificial opal photonic
  crystals and inverse opal structures – fundamentals and applications from
  optics to energy storage}},\ }\href {https://doi.org/10.1039/C5TC01083G}
  {\bibfield  {journal} {\bibinfo  {journal} {J. Mater. Chem. C}\ }\textbf
  {\bibinfo {volume} {3}},\ \bibinfo {pages} {6109} (\bibinfo {year}
  {2015})}\BibitemShut {NoStop}%
\bibitem [{\citenamefont {Krupinski}\ \emph {et~al.}(2019)\citenamefont
  {Krupinski}, \citenamefont {Bali}, \citenamefont {Mitin}, \citenamefont
  {Sobieszczyk}, \citenamefont {Gregor-Pawlowski}, \citenamefont {Zarzycki},
  \citenamefont {Böttger}, \citenamefont {Albrecht}, \citenamefont {Potzger},\
  and\ \citenamefont {Marszalek}}]{Krupinski2019}%
  \BibitemOpen
  \bibfield  {author} {\bibinfo {author} {\bibfnamefont {M.}~\bibnamefont
  {Krupinski}}, \bibinfo {author} {\bibfnamefont {R.}~\bibnamefont {Bali}},
  \bibinfo {author} {\bibfnamefont {D.}~\bibnamefont {Mitin}}, \bibinfo
  {author} {\bibfnamefont {P.}~\bibnamefont {Sobieszczyk}}, \bibinfo {author}
  {\bibfnamefont {J.}~\bibnamefont {Gregor-Pawlowski}}, \bibinfo {author}
  {\bibfnamefont {A.}~\bibnamefont {Zarzycki}}, \bibinfo {author}
  {\bibfnamefont {R.}~\bibnamefont {Böttger}}, \bibinfo {author}
  {\bibfnamefont {M.}~\bibnamefont {Albrecht}}, \bibinfo {author}
  {\bibfnamefont {K.}~\bibnamefont {Potzger}},\ and\ \bibinfo {author}
  {\bibfnamefont {M.}~\bibnamefont {Marszalek}},\ }\bibfield  {title} {\bibinfo
  {title} {{Ion induced ferromagnetism combined with self-assembly for large
  area magnetic modulation of thin films}},\ }\href
  {https://doi.org/10.1039/C8NR10011J} {\bibfield  {journal} {\bibinfo
  {journal} {Nanoscale}\ }\textbf {\bibinfo {volume} {11}},\ \bibinfo {pages}
  {8930} (\bibinfo {year} {2019})}\BibitemShut {NoStop}%
\bibitem [{\citenamefont {Qiu}\ \emph {et~al.}(2022)\citenamefont {Qiu},
  \citenamefont {Akinoglu}, \citenamefont {Luo}, \citenamefont {Konarova},
  \citenamefont {Yun}, \citenamefont {Gentle},\ and\ \citenamefont
  {Wang}}]{Qiu2022}%
  \BibitemOpen
  \bibfield  {author} {\bibinfo {author} {\bibfnamefont {T.}~\bibnamefont
  {Qiu}}, \bibinfo {author} {\bibfnamefont {E.~M.}\ \bibnamefont {Akinoglu}},
  \bibinfo {author} {\bibfnamefont {B.}~\bibnamefont {Luo}}, \bibinfo {author}
  {\bibfnamefont {M.}~\bibnamefont {Konarova}}, \bibinfo {author}
  {\bibfnamefont {J.-H.}\ \bibnamefont {Yun}}, \bibinfo {author} {\bibfnamefont
  {I.~R.}\ \bibnamefont {Gentle}},\ and\ \bibinfo {author} {\bibfnamefont
  {L.}~\bibnamefont {Wang}},\ }\bibfield  {title} {\bibinfo {title}
  {{Nanosphere Lithography: A Versatile Approach to Develop Transparent
  Conductive Films for Optoelectronic Applications}},\ }\href
  {https://doi.org/10.1002/adma.202103842} {\bibfield  {journal} {\bibinfo
  {journal} {Advanced Materials}\ }\textbf {\bibinfo {volume} {34}},\ \bibinfo
  {pages} {2103842} (\bibinfo {year} {2022})}\BibitemShut {NoStop}%
\bibitem [{\citenamefont {Goiriena-Goikoetxea}\ \emph
  {et~al.}(2016)\citenamefont {Goiriena-Goikoetxea}, \citenamefont
  {García-Arribas}, \citenamefont {Rouco}, \citenamefont {Svalov},\ and\
  \citenamefont {Barandiaran}}]{GoirienaGoikoetxea2016}%
  \BibitemOpen
  \bibfield  {author} {\bibinfo {author} {\bibfnamefont {M.}~\bibnamefont
  {Goiriena-Goikoetxea}}, \bibinfo {author} {\bibfnamefont {A.}~\bibnamefont
  {García-Arribas}}, \bibinfo {author} {\bibfnamefont {M.}~\bibnamefont
  {Rouco}}, \bibinfo {author} {\bibfnamefont {A.~V.}\ \bibnamefont {Svalov}},\
  and\ \bibinfo {author} {\bibfnamefont {J.~M.}\ \bibnamefont {Barandiaran}},\
  }\bibfield  {title} {\bibinfo {title} {{High-yield fabrication of 60 nm
  Permalloy nanodiscs in well-defined magnetic vortex state for biomedical
  applications}},\ }\href {https://doi.org/10.1088/0957-4484/27/17/175302}
  {\bibfield  {journal} {\bibinfo  {journal} {Nanotechnology}\ }\textbf
  {\bibinfo {volume} {27}},\ \bibinfo {pages} {175302} (\bibinfo {year}
  {2016})}\BibitemShut {NoStop}%
\bibitem [{\citenamefont {Welbourne}\ \emph {et~al.}(2021)\citenamefont
  {Welbourne}, \citenamefont {Vemulkar},\ and\ \citenamefont
  {Cowburn}}]{Welbourne2021-1}%
  \BibitemOpen
  \bibfield  {author} {\bibinfo {author} {\bibfnamefont {E.~N.}\ \bibnamefont
  {Welbourne}}, \bibinfo {author} {\bibfnamefont {T.}~\bibnamefont
  {Vemulkar}},\ and\ \bibinfo {author} {\bibfnamefont {R.~P.}\ \bibnamefont
  {Cowburn}},\ }\bibfield  {title} {\bibinfo {title} {{High-yield fabrication
  of perpendicularly magnetised synthetic antiferromagnetic nanodiscs}},\
  }\href {https://doi.org/10.1007/s12274-021-3307-1} {\bibfield  {journal}
  {\bibinfo  {journal} {Nano Research}\ }\textbf {\bibinfo {volume} {14}},\
  \bibinfo {pages} {3873} (\bibinfo {year} {2021})}\BibitemShut {NoStop}%
\bibitem [{\citenamefont {Meyer}\ \emph {et~al.}(2021)\citenamefont {Meyer},
  \citenamefont {Bennewitz},\ and\ \citenamefont {Hug}}]{Meyer2021}%
  \BibitemOpen
  \bibfield  {author} {\bibinfo {author} {\bibfnamefont {E.}~\bibnamefont
  {Meyer}}, \bibinfo {author} {\bibfnamefont {R.}~\bibnamefont {Bennewitz}},\
  and\ \bibinfo {author} {\bibfnamefont {H.~J.}\ \bibnamefont {Hug}},\
  }\bibfield  {title} {\bibinfo {title} {{Scanning Probe Microscopy, The Lab on
  a Tip}}\ }\href {https://doi.org/10.1007/978-3-030-37089-3}
  {10.1007/978-3-030-37089-3} (\bibinfo {year} {2021})\BibitemShut {NoStop}%
\bibitem [{\citenamefont {Feng}\ \emph
  {et~al.}(2022{\natexlab{a}})\citenamefont {Feng}, \citenamefont {Vaghefi},
  \citenamefont {Vranjkovic}, \citenamefont {Penedo}, \citenamefont
  {Kappenberger}, \citenamefont {Schwenk}, \citenamefont {Zhao}, \citenamefont
  {Mandru},\ and\ \citenamefont {Hug}}]{Feng2022}%
  \BibitemOpen
  \bibfield  {author} {\bibinfo {author} {\bibfnamefont {Y.}~\bibnamefont
  {Feng}}, \bibinfo {author} {\bibfnamefont {P.~M.}\ \bibnamefont {Vaghefi}},
  \bibinfo {author} {\bibfnamefont {S.}~\bibnamefont {Vranjkovic}}, \bibinfo
  {author} {\bibfnamefont {M.}~\bibnamefont {Penedo}}, \bibinfo {author}
  {\bibfnamefont {P.}~\bibnamefont {Kappenberger}}, \bibinfo {author}
  {\bibfnamefont {J.}~\bibnamefont {Schwenk}}, \bibinfo {author} {\bibfnamefont
  {X.}~\bibnamefont {Zhao}}, \bibinfo {author} {\bibfnamefont {A.-O.}\
  \bibnamefont {Mandru}},\ and\ \bibinfo {author} {\bibfnamefont
  {H.}~\bibnamefont {Hug}},\ }\bibfield  {title} {\bibinfo {title} {{Magnetic
  force microscopy contrast formation and field sensitivity}},\ }\href
  {https://doi.org/10.1016/j.jmmm.2022.169073} {\bibfield  {journal} {\bibinfo
  {journal} {Journal of Magnetism and Magnetic Materials}\ }\textbf {\bibinfo
  {volume} {551}},\ \bibinfo {pages} {169073} (\bibinfo {year}
  {2022}{\natexlab{a}})}\BibitemShut {NoStop}%
\bibitem [{\citenamefont {N{\'e}el}(1962)}]{neel1962new}%
  \BibitemOpen
  \bibfield  {author} {\bibinfo {author} {\bibfnamefont {L.}~\bibnamefont
  {N{\'e}el}},\ }\bibfield  {title} {\bibinfo {title} {Sur un nouveau mode de
  couplage entre les aimantations de deux couches minces ferromagnétiques},\
  }\href@noop {} {\bibfield  {journal} {\bibinfo  {journal} {Comptes Rendus
  Hebdomadaires Des Seances De L Academie Des Sciences}\ }\textbf {\bibinfo
  {volume} {255}},\ \bibinfo {pages} {1676} (\bibinfo {year}
  {1962})}\BibitemShut {NoStop}%
\bibitem [{\citenamefont {Buchholz}\ \emph {et~al.}(2024)\citenamefont
  {Buchholz}, \citenamefont {Sajjamark}, \citenamefont {Franke}, \citenamefont
  {Wei}, \citenamefont {Behrends}, \citenamefont {Münkel}, \citenamefont
  {Grüttner}, \citenamefont {Levan}, \citenamefont {von Elverfeldt},
  \citenamefont {Graeser}, \citenamefont {Buzug}, \citenamefont {Bär},\ and\
  \citenamefont {Hofmann}}]{thno86759}%
  \BibitemOpen
  \bibfield  {author} {\bibinfo {author} {\bibfnamefont {O.}~\bibnamefont
  {Buchholz}}, \bibinfo {author} {\bibfnamefont {K.}~\bibnamefont {Sajjamark}},
  \bibinfo {author} {\bibfnamefont {J.}~\bibnamefont {Franke}}, \bibinfo
  {author} {\bibfnamefont {H.}~\bibnamefont {Wei}}, \bibinfo {author}
  {\bibfnamefont {A.}~\bibnamefont {Behrends}}, \bibinfo {author}
  {\bibfnamefont {C.}~\bibnamefont {Münkel}}, \bibinfo {author} {\bibfnamefont
  {C.}~\bibnamefont {Grüttner}}, \bibinfo {author} {\bibfnamefont
  {P.}~\bibnamefont {Levan}}, \bibinfo {author} {\bibfnamefont
  {D.}~\bibnamefont {von Elverfeldt}}, \bibinfo {author} {\bibfnamefont
  {M.}~\bibnamefont {Graeser}}, \bibinfo {author} {\bibfnamefont
  {T.}~\bibnamefont {Buzug}}, \bibinfo {author} {\bibfnamefont
  {S.}~\bibnamefont {Bär}},\ and\ \bibinfo {author} {\bibfnamefont {U.~G.}\
  \bibnamefont {Hofmann}},\ }\bibfield  {title} {\bibinfo {title} {<i>in
  situ</i> theranostic platform combining highly localized magnetic fluid
  hyperthermia, magnetic particle imaging, and thermometry in 3d},\ }\href
  {https://doi.org/10.7150/thno.86759} {\bibfield  {journal} {\bibinfo
  {journal} {Theranostics}\ }\textbf {\bibinfo {volume} {14}},\ \bibinfo
  {pages} {324} (\bibinfo {year} {2024})}\BibitemShut {NoStop}%
\bibitem [{\citenamefont {Vogel}\ \emph {et~al.}(2021)\citenamefont {Vogel},
  \citenamefont {Kampf}, \citenamefont {Rückert}, \citenamefont {Grüttner},
  \citenamefont {Kowalski}, \citenamefont {Teller},\ and\ \citenamefont
  {Behr}}]{SynomagVogel}%
  \BibitemOpen
  \bibfield  {author} {\bibinfo {author} {\bibfnamefont {P.}~\bibnamefont
  {Vogel}}, \bibinfo {author} {\bibfnamefont {T.}~\bibnamefont {Kampf}},
  \bibinfo {author} {\bibfnamefont {M.}~\bibnamefont {Rückert}}, \bibinfo
  {author} {\bibfnamefont {C.}~\bibnamefont {Grüttner}}, \bibinfo {author}
  {\bibfnamefont {A.}~\bibnamefont {Kowalski}}, \bibinfo {author}
  {\bibfnamefont {H.}~\bibnamefont {Teller}},\ and\ \bibinfo {author}
  {\bibfnamefont {V.}~\bibnamefont {Behr}},\ }\bibfield  {title} {\bibinfo
  {title} {Synomag®: The new high-performance tracer for magnetic particle
  imaging},\ }\href@noop {} {\bibfield  {journal} {\bibinfo  {journal} {Int. J.
  Magn. Part. Imag.}\ }\textbf {\bibinfo {volume} {7}},\ \bibinfo {pages}
  {2103003} (\bibinfo {year} {2021})}\BibitemShut {NoStop}%
\bibitem [{\citenamefont {Bruckner}\ \emph {et~al.}(2023)\citenamefont
  {Bruckner}, \citenamefont {Koraltan}, \citenamefont {Abert},\ and\
  \citenamefont {Suess}}]{bruckner_magnumnp_2023}%
  \BibitemOpen
  \bibfield  {author} {\bibinfo {author} {\bibfnamefont {F.}~\bibnamefont
  {Bruckner}}, \bibinfo {author} {\bibfnamefont {S.}~\bibnamefont {Koraltan}},
  \bibinfo {author} {\bibfnamefont {C.}~\bibnamefont {Abert}},\ and\ \bibinfo
  {author} {\bibfnamefont {D.}~\bibnamefont {Suess}},\ }\bibfield  {title}
  {\bibinfo {title} {magnum.np: a {PyTorch} based {GPU} enhanced finite
  difference micromagnetic simulation framework for high level development and
  inverse design},\ }\href {https://doi.org/10.1038/s41598-023-39192-5}
  {\bibfield  {journal} {\bibinfo  {journal} {Scientific Reports}\ }\textbf
  {\bibinfo {volume} {13}},\ \bibinfo {pages} {12054} (\bibinfo {year}
  {2023})}\BibitemShut {NoStop}%
\bibitem [{\citenamefont {Landau}\ and\ \citenamefont
  {Lifshitz}(1935)}]{landau_theory_1935}%
  \BibitemOpen
  \bibfield  {author} {\bibinfo {author} {\bibfnamefont {L.~D.}\ \bibnamefont
  {Landau}}\ and\ \bibinfo {author} {\bibfnamefont {E.~M.}\ \bibnamefont
  {Lifshitz}},\ }\bibfield  {title} {\bibinfo {title} {Theory of the dispersion
  of magnetic permeability in ferromagnetic bodies},\ }\href@noop {} {\bibfield
   {journal} {\bibinfo  {journal} {Phys. Z. Sowjetunion}\ }\textbf {\bibinfo
  {volume} {8}},\ \bibinfo {pages} {153} (\bibinfo {year} {1935})}\BibitemShut
  {NoStop}%
\bibitem [{\citenamefont {Gilbert}(1955)}]{gilbert_lagrangian_1955}%
  \BibitemOpen
  \bibfield  {author} {\bibinfo {author} {\bibfnamefont {T.~L.}\ \bibnamefont
  {Gilbert}},\ }\bibfield  {title} {\bibinfo {title} {A {Lagrangian}
  formulation of the gyromagnetic equation of the magnetization field},\
  }\href@noop {} {\bibfield  {journal} {\bibinfo  {journal} {Phys. Rev.}\
  }\textbf {\bibinfo {volume} {100}},\ \bibinfo {pages} {1243} (\bibinfo {year}
  {1955})}\BibitemShut {NoStop}%
\bibitem [{\citenamefont {Gilbert}(2004)}]{gilbert_phenomenological_2004}%
  \BibitemOpen
  \bibfield  {author} {\bibinfo {author} {\bibfnamefont {T.~L.}\ \bibnamefont
  {Gilbert}},\ }\bibfield  {title} {\bibinfo {title} {A phenomenological theory
  of damping in ferromagnetic materials},\ }\href@noop {} {\bibfield  {journal}
  {\bibinfo  {journal} {IEEE transactions on magnetics}\ }\textbf {\bibinfo
  {volume} {40}},\ \bibinfo {pages} {3443} (\bibinfo {year} {2004})},\ \bibinfo
  {note} {publisher: IEEE}\BibitemShut {NoStop}%
\bibitem [{\citenamefont {Schendel}\ \emph {et~al.}(2000)\citenamefont
  {Schendel}, \citenamefont {Hug}, \citenamefont {Stiefel}, \citenamefont
  {Martin},\ and\ \citenamefont {untherodt}}]{VanSchendel2000}%
  \BibitemOpen
  \bibfield  {author} {\bibinfo {author} {\bibfnamefont {P.~J. A.~v.}\
  \bibnamefont {Schendel}}, \bibinfo {author} {\bibfnamefont {H.~J.}\
  \bibnamefont {Hug}}, \bibinfo {author} {\bibfnamefont {B.}~\bibnamefont
  {Stiefel}}, \bibinfo {author} {\bibfnamefont {S.}~\bibnamefont {Martin}},\
  and\ \bibinfo {author} {\bibfnamefont {H.~J.~G.}\ \bibnamefont {untherodt}},\
  }\bibfield  {title} {\bibinfo {title} {{A method for the calibration of
  magnetic force microscopy tips}},\ }\href {https://doi.org/10.1063/1.373678}
  {\bibfield  {journal} {\bibinfo  {journal} {Journal Of Applied Physics}\
  }\textbf {\bibinfo {volume} {88}},\ \bibinfo {pages} {435 } (\bibinfo {year}
  {2000})}\BibitemShut {NoStop}%
\bibitem [{\citenamefont {Feng}\ \emph
  {et~al.}(2022{\natexlab{b}})\citenamefont {Feng}, \citenamefont {Mandru},
  \citenamefont {Yıldırım},\ and\ \citenamefont {Hug}}]{Feng2022-2}%
  \BibitemOpen
  \bibfield  {author} {\bibinfo {author} {\bibfnamefont {Y.}~\bibnamefont
  {Feng}}, \bibinfo {author} {\bibfnamefont {A.-O.}\ \bibnamefont {Mandru}},
  \bibinfo {author} {\bibfnamefont {O.}~\bibnamefont {Yıldırım}},\ and\
  \bibinfo {author} {\bibfnamefont {H.}~\bibnamefont {Hug}},\ }\bibfield
  {title} {\bibinfo {title} {{Quantitative Magnetic Force Microscopy:
  Transfer-Function Method Revisited}},\ }\href
  {https://doi.org/10.1103/physrevapplied.18.024016} {\bibfield  {journal}
  {\bibinfo  {journal} {Physical Review Applied}\ }\textbf {\bibinfo {volume}
  {18}},\ \bibinfo {pages} {024016} (\bibinfo {year}
  {2022}{\natexlab{b}})}\BibitemShut {NoStop}%
\bibitem [{\citenamefont {Sherwood}(2012)}]{sherwood_resistance_2012}%
  \BibitemOpen
  \bibfield  {author} {\bibinfo {author} {\bibfnamefont {J.~D.}\ \bibnamefont
  {Sherwood}},\ }\bibfield  {title} {\bibinfo {title} {Resistance coefficients
  for {Stokes} flow around a disk with a {Navier} slip condition},\ }\href
  {https://doi.org/10.1063/1.4754869} {\bibfield  {journal} {\bibinfo
  {journal} {Physics of Fluids}\ }\textbf {\bibinfo {volume} {24}},\ \bibinfo
  {pages} {093103} (\bibinfo {year} {2012})}\BibitemShut {NoStop}%
\bibitem [{\citenamefont {Zhao}\ \emph {et~al.}(2018)\citenamefont {Zhao},
  \citenamefont {Schwenk}, \citenamefont {Mandru}, \citenamefont {Penedo},
  \citenamefont {Bacani}, \citenamefont {Marioni},\ and\ \citenamefont
  {Hug}}]{Zhao2018}%
  \BibitemOpen
  \bibfield  {author} {\bibinfo {author} {\bibfnamefont {X.}~\bibnamefont
  {Zhao}}, \bibinfo {author} {\bibfnamefont {J.}~\bibnamefont {Schwenk}},
  \bibinfo {author} {\bibfnamefont {A.~O.}\ \bibnamefont {Mandru}}, \bibinfo
  {author} {\bibfnamefont {M.}~\bibnamefont {Penedo}}, \bibinfo {author}
  {\bibfnamefont {M.}~\bibnamefont {Bacani}}, \bibinfo {author} {\bibfnamefont
  {M.~A.}\ \bibnamefont {Marioni}},\ and\ \bibinfo {author} {\bibfnamefont
  {H.~J.}\ \bibnamefont {Hug}},\ }\bibfield  {title} {\bibinfo {title}
  {{Magnetic force microscopy with frequency-modulated capacitive tip–sample
  distance control}},\ }\href {https://doi.org/10.1088/1367-2630/aa9ca9}
  {\bibfield  {journal} {\bibinfo  {journal} {New Journal of Physics}\ }\textbf
  {\bibinfo {volume} {20}},\ \bibinfo {pages} {013018 } (\bibinfo {year}
  {2018})}\BibitemShut {NoStop}%
\end{thebibliography}%

\noindent\textbf{Acknowledgements}\\
We acknowledge the funding of S. Scheibler's Ph.D. thesis work, through an Empa internal research projects, and an additional year as a postdoc by Empa's entrepreneural fellowship program. I.K.H. acknowledges support by the Swiss National Science Foundation SNSF Eccellenza program (grant no. 181290). \\

\noindent\textbf{Author contributions}\\
H.~J.~H, D.~S., and I.~K.~H conceived the project. S.~S. deposited the magnetic multilayer films, and performed the magnetometry experiments and data analysis. The polystyrene sphere self-assembly and successive oxygen plasma etching to reduce the PS sphere diameters was performed by M.~K., while S.~S. performed the successive ion etching to pattern the SAF-MDP islands and successively release these from the wafer support to form the SAF-MDP suspensions. The MFM and data acquisition and processing was performed by H.~J.~H and R.~P.-P. The micromagnetic and microfluidic modeling was performed by S.~H., S.~K., H.~J.~H, and D.~S. The hyperthermia apparatus was designed and setup by H.~W., J.~A., and M.~G., while the corresponding experiments and data analysis were performed by H.~W., S.~S., H.~J.~H, and M.~G.. The manuscript was conceived by
H.~J.~H, D.~S., and I.~K.~H., and all authors discussed and contributed to the final manuscript.
\\

\noindent\textbf{Competing interests}\\
Nature Journals require authors to declare any competing interests in relation to the work described. Information on this policy is available \href{http://www.nature.com/authors/policies/competing.html}{here}. \\

\section*{Supplementary information}
\subsection*{Magnetic thin film multilayer deposition}
The multilayers, including a 7\,nm thick top sacrificial layer of MgO, were sputter-deposited onto a 2-inch Si wafer using an AJA DC/RF magnetron sputtering system under a working pressure of $10^{-3}$\,Pa of Ar gas and a base pressure below $1 \times 10^{-8}$\,Pa. To establish a well-defined uniaxial in-plane magnetic anisotropy, the substrate was fixed to a custom-built sample holder containing two NdFe permanent magnets, generating a uniform field of 400\,Oe within the substrate plane.

\subsection*{Modeling Work}
\subsubsection*{Micromagnetic Simulations} 
The micromagnetic simulations were performed with the python library \texttt{magnum.np} \cite{bruckner_magnumnp_2023} using a finite difference method with a regular rectangular mesh. The real dimensions of the simulation box around the disk are $500\,{\rm nm} \times 500\,{\rm nm} \times 16\,$nm and is discretized with rectangular cells. Each layer is 6.3\,nm thick and separated by a 3.4\,nm thick spacer layer. This results in a simulation box of $100 \times 100 \times 3$ cells, with a discretization of $5\,{\rm nm} \times 5\,{\rm nm} \times 6.3\,$nm, while the cells in the spacer layer have a size of $5\,{\rm nm} \times 5\,{\rm nm} \times 3.4\,$nm.
The magnetic parameters in the spacer layer and the material outside of the disk radius are set to zero, e.g. $M_s = 0$.
The micromagnetic framework solves the Landau-Lifshitz-Gilbert (LLG) equation \cite{landau_theory_1935,gilbert_lagrangian_1955,gilbert_phenomenological_2004} in each cell, which describes the change of the magnetic moment according to all contributions to the effective field.
\begin{equation}
    \mathbf{\dot{m}} = - \frac{\mu_0 \gamma}{1 + \alpha^2} \mathbf{m} \times \mathbf{H}_\mathrm{eff} - \frac{\alpha \mu_0 \gamma}{1 + \alpha^2} \; \mathbf{m} \times (\mathbf{m} \times \mathbf{H}_\mathrm{eff}) \,\, .
\label{eq:LLG}
\end{equation}

Here, $m$ is the unit vector pointing in the direction of the magnetic moment, $\mu_0$ is the vacuum permeability constant, $\gamma$ is the gyromagnetic ratio, $\alpha$ is the Gilbert damping parameter, and $\mathbf{H}_\mathrm{eff}$ is the effective field. The first term describes the precession and the second term the damping of the magnetic moment.
Each layer simulated with in-plane anisotropy, exchange coupling and also included is the demagnetization energy which is mainly responsible for the magnetic coupling of the two layers via the stray field. Additionally, Ruderman-Kittel-Kasuya-Yosida (RKKY) interactions are included to emulate the orange-peel effect by introducing a small alignment field. Lastly, an external field is applied in the simulations. This results in an effective field consisting of the anisotropy field, the exchange field in each layer, the demagnetization field, the RKKY field and the externally applied field. In order to avoid any numerical instabilities due to symmetry, the field is applied at an angle of 5 degrees relative to the easy axis and at an angle of 89 degrees for the hard axis simulations. The saturation magnetization in one layer is increased by 2 \% to further lower the symmetry of the system.
For the measurements of the $M(H)$-loops (Fig.\,\ref{Fig:micromagmodel}{\bf a} and Fig.\,\ref{Fig:results}{\bf j}) and the simulated magnetic  moment distributions and MFM data (Fig.\,\ref{Fig:results}{\bf m} to {\bf p}) the applied field was cycled from -100\,mT to 100\,mT changing by 2\,mT each \SI{1e-9}{\second} and finishing one whole cycle in 200\,ns.

\subsubsection*{Modeling of Magnetic Force Microscopy Contrast}
In order to simulate the MFM frequency shift contrast at a given scanning height of 20\,nm, we consider a magnetization state as our input state which was obtained by numerically solving the LLG, as explained in the previous paragraph. We increase the height of the simulation box so that the center of the last cell corresponds to the desired MFM scanning height, where all magnetic parameters are set to zero, e.g. $M_{\rm s} = 0$, thus mimicking vacuum. The stray field $\mathbf{H}$ originating from the main magnetic body are now calculated in the large box. For the modeling of the MFM contrast, the MFM tip is assumed to be a point dipole. Note that MFM tips could be calibrated\,\cite{VanSchendel2000,Feng2022-2}. Then the calibrated tip response could be integrated into the model to obtain a more quantitative comparison of the modeled with the measured MFM contrast. This was however not the focus of our work here because without a correct spatial description of anisotropy variations a quantitative agreement can never be obtained.

To simulate magnetic materials in a more realistic way, we assume a distribution of anisotropy values and direction by using the built-in Voronoi tesselation in \texttt{magnum.np}. To do so, we first calculate an initial Voronoi tessellation from a randomly distributed cloud of points. The cores of the Voronoi cells are then used as an input, and we iteratively smooth the corners of our Voronoi tessellation to simulate the real grains, and grain bonds of the magnetic materials used in this study. The average grain size of the Voronoi cells is about 20 nm. In the SAF layer, we assume that each magnetic layer has its own distribution.
The anisotropy energy constant in each grain is pulled from a Gaussian distribution with mean = 20\,kJ m$^{-3}$ and a standard deviation of 2\,kJ m$^{-3}$. Additionally, the easy axis is pulled from a uniform distribution between -3 and +3 degrees relative to the x-axis of the simulation grid leading to a variation of the easy axis in the x-y plane.

\subsubsection*{Mechanofluidic modeling.} 
Similarly to the full micromagnetic simulations, these macrospin simulations solve the LLG but only for two exchange coupled spins located at the center of a virtual disk. Different from the micromagnetic simulations, the macrospin simulations allow for an additional degree of freedom by letting the disk rotate according to the equations of motion derived from the torques acting on the disk and the conservation of angular momentum of the system. Based on previous work with spherical particles \cite{Helbig2023}, the viscous torque and inertia were adapted for a disk.

The viscous torque $\mathbf{\tau_{visc}}$ with a no-slip condition \cite{sherwood_resistance_2012} and the inertial torque of a disk $\mathbf{L}_{inert}$ are given by the expressions
\begin{align}\label{eq:tauvisc}
    \mathbf{\tau_{visc}} = -\dfrac{32}{3} \; \eta r^3 \mathbf{\dot{\phi}} \; \; , &&
	\mathbf{L}_{inert} = \dfrac{1}{12} \mathfrak{m} \mathbf{\dot{\phi}} \left[ \begin{matrix}
		3 \, r^2 + h^2 \\
		3 \, r^2 + h^2 \\
		6 \, r^2
	\end{matrix} \right] ,
\end{align}
where $\eta$ denotes the dynamic viscosity of the carrier fluid (here assumed to be $\eta = \SI{0.89e-3}{\pascal \second}$ similar to water at 25$^\circ$C), $r$ is the radius of the disk, and $\mathbf{\dot{\phi}}$ is the angular velocity of the disk rotating about the angle $\phi$. In the inertia tensor $\mathfrak{m}$ denotes the mass of the particle, and $h$ is the height of the disk.

The magnetic contributions, namely the spin angular momentum $\mathbf{L}_\mathrm{spin}$ and the torque $\mathbf{\tau}_\mathrm{mag}$ exerted by the external field $\mathbf{H}_\mathrm{ext}$, stay the same as with the spherical particles, since they are not influenced by the change of shape:
\begin{align}\label{eq:Lspin}
	\mathbf{L}_\mathrm{spin} = - \frac{M_s V_m}{\gamma}  \mathbf{m} \; \; , &&
	\mathbf{\tau}_\mathrm{mag} = \mu_0 M_\mathrm{s} V_\mathrm{m} \mathbf{m} \times \mathbf{H}_\mathrm{ext} .
\end{align}
Here, $M_{\rm s}$ denotes the saturation magnetization, and $V_{\rm m}$ is the total volume of the magnetic material.
The conservation of angular momentum requires that:
\begin{equation}\label{eq:conservation}
		\mathbf{\dot{L}}_\mathrm{spin}+\mathbf{\dot{L}}_\mathrm{inert}=\mathbf{\tau}_\mathrm{mag}+\mathbf{\tau}_\mathrm{visc}
\end{equation}
Combining these equations into a self-consistent solution and rearranging the terms to receive an update scheme for the rotation of the particle's easy axis yields:
\begin{equation}\label{eq:eom}
	\mathbf{\ddot{\phi}} = \frac{12}{\mathfrak{m}} \left ( \mu_0 M_\mathrm{s} V_\mathrm{m} \mathbf{m} \times \mathbf{H}_\mathrm{ext} - \dfrac{32}{3} \; \eta r^3 \mathbf{\dot{\phi}} + \frac{M_s V_m}{\gamma}  \mathbf{\dot{m}} \right ) \left[ \begin{matrix}
		\frac{1}{3 \, r^2 + h^2} \\
		\frac{1}{3 \, r^2 + h^2} \\
		\frac{1}{6 \, r^2}
	\end{matrix} \right].
\end{equation}
In the case of the macrospin model, the effective field for the LLG consists of the anisotropy field, the external field and an antiferromagnetic exchange field.

\subsection*{PS sphere self-assembly}
Close-packed arrays of magnetic disks were fabricated using nanosphere lithography. At the beginning of this process, the wafers with magnetic thin film multilayers were treated in soft nitrogen-oxygen plasma for 5 min in order to remove contaminations and to enhance their wettability. Shortly after the plasma treatment, the wafers were submerged under surface of distilled water inside 15 cm Petri dish, where self-assembly process was carried out. Monodisperse aqueous suspension of non-functionalized and uncrosslinked polystyrene (PS) particles with average diameter of 658 nm (standard deviation of 15 nm) from MicroParticles GmbH Berlin was diluted by mixing with equal volume of ethanol (99,5\% pure). The suspension was applied to the surface of the water by glass Pasteur pipette with a rate of approximately 1 µl/s. When the entire surface of the water in the Petri dish has been filled with PS particles, 2D crystallization took place spontaneously leading to a highly ordered hexagonal close-packed monolayer of the particles. The monolayers were deposited on the magnetic multilayers by slow water evaporation at room temperature. Next, RF-plasma etching was used (MiniFlecto, Plasma Technology GmbH, Herrenberg, Germany), resulting in the decrease of the spheres size, but maintaining their original positions and arrangement. The plasma process was performed in oxygen and argon atmosphere under pressure of 0.15 mbar and temperature of 30 °C. During the process, a continuous flow of oxygen and argon was applied to 2 sccm and 1 sccm, respectively. After 510 s of the etching, the PS particles diameter decreased to 500 nm, which was confirmed by inspection under a scanning electron microscope.

\subsection*{Suspension Fabrication and Characterization}
The sample concentration and elemental composition of the SAF MDPs in water was analyzed via inductively coupled plasma mass spectrometry (ICP-MS). Samples with a volume of \SI{50}{\micro\liter} were digested in quartz tubes using a mixture of hydrochloric acid (37\%, Normatom, VWR) and nitric acid (69\%, Normatom, VWR) in a pressurized microwave system (TurboWAVE, MLS GmbH, Germany) at \SI{230}{\degreeCelsius} and \SI{120}{bar} for 19 minutes. After digestion, the samples were transferred to \SI{50}{ml} Falcon tubes and diluted with ultrapure water. Analysis for Al, Co, Zr, and Sm was conducted using a 7900 single-quadrupole ICP-MS (Agilent Technologies, CA). Isotopes \(^{27}\)Al were measured in No-gas mode, while \(^{59}\)Co, \(^{90}\)Zr, and \(^{147}\)Sm were determined in He-collision mode. Non-spectral interferences were corrected using an internal standard solution containing Li (for Al correction) and Rh (for all other elements), which was mixed online with the sample. Calibration was performed with certified element standards (Inorganic ventures) diluted in the same acid matrix as the samples.

\subsection*{High-resolution Magnetic Force Microscopy with applied in-plane fields}
Performing magnetic force microscopy with high spatial resolution with negligible perturbation of the micromagnetic state of the SAF-MPDs still attached to the wafer remains a challenging experiment, requiring both specialized MFM instrumentation as well as operation modes and data processing (see chapter 4 in ref.\,\cite{Meyer2021}). 

Here we used a home-built magnetic force microscope operating in vacuum, using an SS-ISC cantilever from Team Nanotech without any coating. The tip was made sensitive to magnetic fields by sputter-deposition of a 
Ta(2\,nm)/Co(6\,nm)/Ta(4\,nm) seed, magnetic and oxidation protection layer system. The thicknesses are nominal thicknesses obtained for a substrate placed perpendicular to the incoming sputtered particles. 
To achieve a high quality factor during vacuum operation, the cantilever base was masked\,\cite{Feng2022} to avoid coating the parts of the cantilever near the chip experiencing the highest strain upon cantilever deflection. The free resonance frequency $f_0 = 55.09897\,$kHz and quality factor $Q=237'192$ were then found by sweeping the excitation frequency through through the cantilever resonance. The cantilever stiffness $c = 1.12\,$N/m is obtained from the known materials constants of silicon and the cantilever's length and width\,\cite{Feng2022}. The relatively low cantilever stiffness and high quality factor then provide a measurement sensitivity (for the chosen oscillation amplitude of $A_{\rm rms}=5\,$nm) of $\sim 0.1\mu{\rm N}/\sqrt{\rm Hz}$ that is about a factor of 40 above that typically obtained with MFM instruments operated with conventional cantilevers under ambient conditions\,\cite{Feng2022}. This enhanced measurement sensitivity permits the use of ultra-low magnetic moment tips which have only a small stray field and thus do not noticeably influence the micromagnetic state of the sample such that the true micromagnetic state of the SAF-MDPs can be observed. 

The cantilever was driven on resonance using a phase-locked loop with an oscillation amplitude kept constant at $A_{\rm rms}=5\,$nm. The measured quantity then is the shift of the resonance frequency away from the free cantilever resonance frequency. 
The MFM data acquisition was performed with the tip scanning parallel to the average slope of the sample. The distance between the tip and the nearly flat tops of the SAF-MDP islands was kept constant in average using our capacitive frequency modulated distance control operation mode\,\cite{Zhao2018}.

Generally forces of different physical nature are simultaneously acting on the tip of a scanning (or magnetic) force microscope and thus contribute to the measured contrast. Sophisticated differential imaging techniques were hence applied\,\cite{Feng2022,Meyer2021} to disentangle the different contrast contributions and ultimately obtain the data shown in Figs.\,\ref{Fig:results}{\bf b} to {\bf i}.

In-plane fields between $\pm 40\,mT$ were applied to the sample by means of a permanent magnet position with an in-vacuum piezo motor linear actuator. The sign of the field was set by a rotation of the magnet such that its north or south pole was facing the sample. 

\subsection*{Hyperthermia Apparatus and Experiments}
The hyperthermia device consists of a transmission coil build of a copper hollow-conductor generating a field of $40\,$mT at $254\,$A current. In order to generate such high ac currents an oszillating circuit is used. This circuit provides reflective free impedance matching to the $50\,\Omega$ desired load for the high frequency amplifier. As frequency $305\,$kHz was chosen. At full field strength the voltage on the coil reach values above $700\,$V.

\end{document}